%% file: ArXiv.tex
\documentclass[reprint,amsmath,amssymb,aps,prl,floatfix,superscriptaddress,twocolumn]{revtex4-2}

\input{definitions}

\newcommand{\env}[1]{\texttt{#1}}
\usepackage{float}
\usepackage{graphicx}
\usepackage{lineno}
\usepackage{docs}%
\usepackage{pifont}
\usepackage{xcolor} 
\usepackage{amsmath}
\usepackage{amsfonts}
\usepackage{comment}
\usepackage[colorlinks=true,linkcolor=blue]{hyperref}
\usepackage[capitalize]{cleveref}

\usepackage{float}
\usepackage[normalem]{ulem}

\begin{document}


\title{Hydrodynamic capture and release of a microswimmer by a meniscus corner}
\author{Subhasish Guchhait}%

\affiliation{
 Department of Mechanical and Aerospace Engineering,\\
Indian Institute of Technology Hyderabad,
Kandi, Sangareddy, Telangana 502285, India }
\author{Harshita Tiwari}%
 \altaffiliation{Present address: Chevron ENGINE, Bengaluru, India}
\affiliation{
 Department of Chemical Engineering,\\
 Indian Institute of Technology, Madras, Tamil Nadu 600036, India}

\author{Sumesh P. Thampi}%
\email{sumesh@iitm.ac.in }
\affiliation{
 Department of Chemical Engineering,\\
 Indian Institute of Technology, Madras, Tamil Nadu 600036, India}
\author{Ranabir Dey}%
\email{ranabir@mae.iith.ac.in}
\affiliation{ 
Department of Mechanical and Aerospace Engineering,\\
Indian Institute of Technology Hyderabad,
Kandi, Sangareddy, Telangana 502285, India }
%

\begin{abstract}
Biological microswimmers alter their motility in complex corner geometries, facilitating their survival.
However, the dynamical features of low-Reynolds-number swimming at corners remain undefined.
Here, we use active droplet microswimmers near a confined meniscus in a microchannel as a model system to study how microswimmer-corner interactions determine swimming patterns.
Combining experiments, theory and simulations, we show that pusher-type micrsowimmers are attracted towards a meniscus corner, followed by transient trapping and eventual escape.
We demonstrate that hydrodynamic interactions with the wall-interface corner intimately dictate the attraction and trapping or escape of the microswimmer on the basis of its  strength. 
We show that the swimming trajectory at the meniscus corner can be tuned depending on the type of the microswimmer, the corner geometry and the viscosity ratio for the liquid interface.
Our study provides a simple way to manipulate microswimmers by exploiting their hydrodynamic interactions near corner geometries.
\end{abstract}
\maketitle
Low-Reynolds-number biological microswimmers including bacteria, algae, and sperm cells inhabit complex geometrical environments where they interact with boundaries such as corners \cite{Sperm_corner1,Sperm_corner2,Bacteria_corner_1,Bacteria_corner_2}, folds or pores \cite{Ecoli_fold,Ecoli_pore,PRE_ECOLI_fold}, interfaces \cite{Bacteria_interface,sperm_interface}, and contact lines \cite{Bacterial_contact_line,Algae_contact_line}.
In such environments, microswimmers can alter their locomotion \cite{sperm_interface}, resulting in geometry-guided motion \cite{Bacteria_corner_1,Bacteria_corner_2,Sperm_corner1,Sperm_corner2} or corner accumulation \cite{Bacterial_contact_line,Algae_contact_line,Ecoli_fold,Ecoli_pore}. 
Evolutionarily, these environments may facilitate survival via oxygen transport \cite{Bacterial_contact_line}, nutrient uptake \cite{Bacterial_contact_line,Algae_contact_line}, and biofilm formation \cite{marra2025microfluidics,krsmanovic2021hydrodynamics}. 
For artificial microswimmers, however, geometrical complexity, like corners, poses an operational hindrance for applications such as targeted drug-delivery, as in branching microvascular networks, leaky vasculature in tumours, and mucus-lined pathways \cite{medina2018micro, lin2021self}. 
While the swimming characteristics depend both on the nature of the microswimmer (pusher, puller, or neutral swimmer) and the nearby boundary (e.g. flat or curved walls and fluid interfaces) \cite{berke2008hydrodynamic, Bacteria_interface, kantsler2013ciliary,bolitho2020periodic,htet2024hydrodynamic}, the generic dynamical features that emerge in wall-interface or wall-wall corner geometries—ubiquitous geometric elements of natural and engineered environments—have not yet been identified.

Theoretical and numerical studies have established how hydrodynamic interactions generated by a self-propelling swimmer cause it to approach, traverse or repel from nearby walls \cite{lauga2006swimming,JFM_Lauga_2012, ishimoto2013squirmer, Byjesh_PRF} or fluid interfaces \cite{JFM_Lauga_2012,POF_Lauga_2014, 2018_soft_matter_near_drop, mishra2026interface}.
In corner geometries formed by idealized free-slip interfaces, the attraction to or repulsion from the corner apex depends on the interplay between the type of swimmer and the opening angle of the corner \cite{2023_POF_corner}.
Numerical studies have also shown the segregation or accumulation of active particles due to interaction with wedge-shaped solid obstacles \cite{PRL_model_ecoli_corner, 2018_soft_matter_trapped_corner, 2020_scientific_reports_trapped_corner, 2025_PRE_trapped_corner}. 
However, the role of hydrodynamics in determining microswimming patterns near corner geometries remains unclear. 
This is in part because corners involve simultaneous hydrodynamic interactions between microswimmers and multiple boundaries, often of different types-solid walls and liquid interfaces. 

Artificial microswimmers, like Janus particles that self-propel via phoretic effects \cite{self-diffusiophoresis, howse2007self, moran2017phoretic} and active droplets driven by Marangoni stresses \cite{Swimmingdroplets2016, michelin2023_selfpropulsion}, are common agents for autonomous cargo-delivery in complex environments. 
Even for such synthetic microswimmers, experimental studies commonly relate to their dynamics near solid walls \cite{das2015boundaries, simmchen2016topographical, de2019flow, ketzetzi2020slip, ketzetzi2020diffusion}, curved boundaries \cite{jin2019fine}, or liquid-liquid/air interfaces \cite{wang2015enhanced, 2019_soft_matter_near_interface, Langmuir_2020_interface}.
Generally, artificial microswimmers tend to reorient and swim along solid walls \cite{simmchen2016topographical}, interfaces and contact lines \cite{2019_soft_matter_near_interface, Langmuir_2020_interface}, demonstrating geometry-guided motion.
However, the swimming patterns at complex corner geometries, especially those formed by a meniscus in a strong confinement, remain unknown despite their common occurrence in microfluidic and biomedical applications.   

Here, we combine experiments, theory, and simulations to investigate microswimming near complex corner geometries: experiments on self-propelling active droplets near liquid-liquid/air meniscus corners formed in rigid microchannels, theoretical analysis using the method of images, and hybrid lattice–Boltzmann simulations.
Our results show that microswimmers are attracted towards the corner apex, and are temporarily trapped before escaping.
We show that while the type and the strength of the microswimmer (force dipole) initiate the attraction and eventually dictates the trapping, the finite size of the microswimmer (source dipole) promotes escape.
The latter process becomes difficult for increasing swimmer strength and smaller corner opening angles. 
We also present phase diagrams of all possible microswimming trajectories depending on the microswimmer type, corner geometry, and the viscosity ratio for the meniscus.

\begin{figure}[h!]
\centering
 \includegraphics[width=8.6cm]{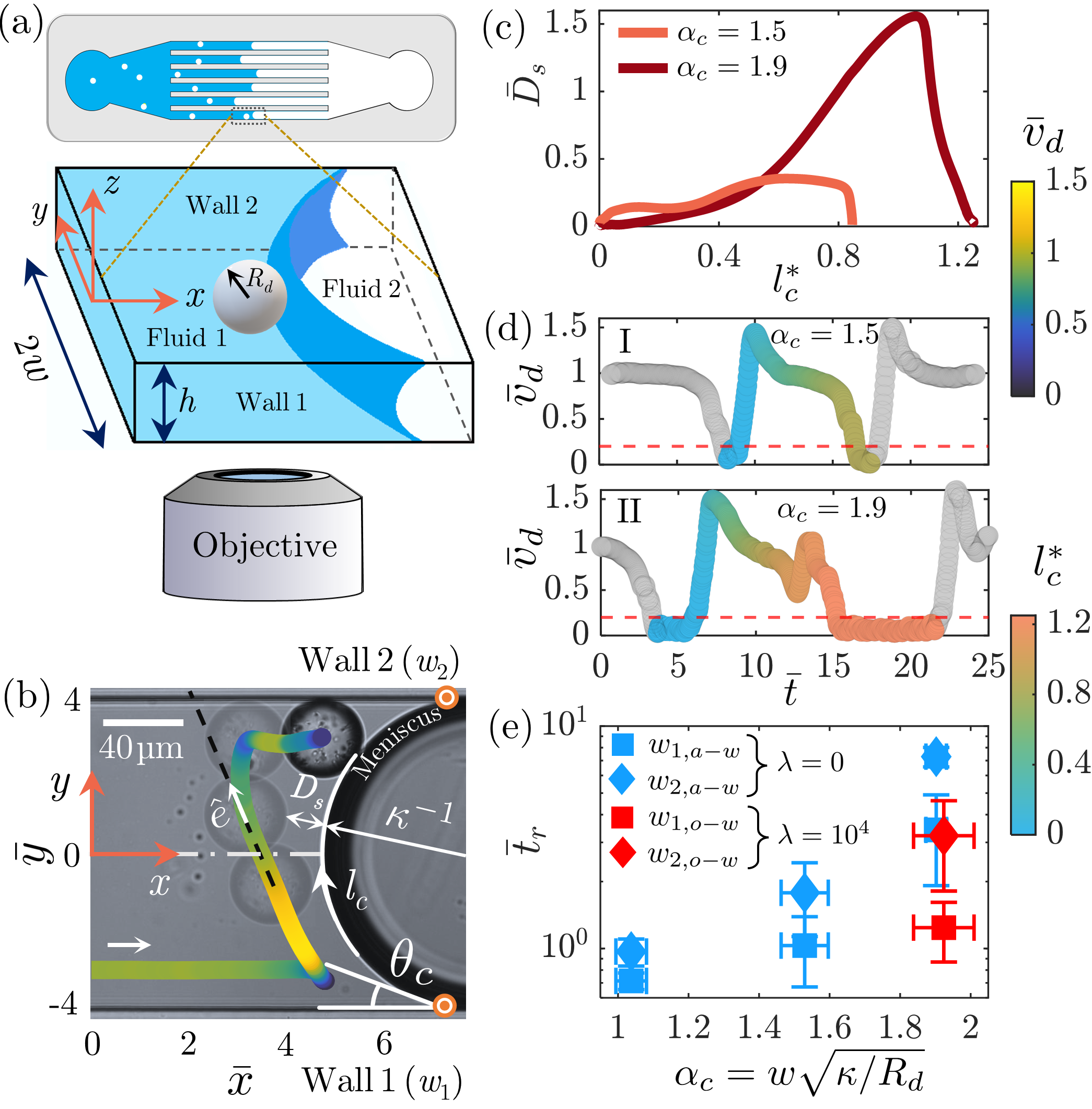}
 \caption{\label{fgr:1}Dynamics of a self-propelling active droplet near a meniscus inside a microchannel. 
 (a) Schematic of the experimental setup. 
 (b) A time-lapse microscopy image showing the swimming trajectory of the active droplet along a confined surfactant solution-air meniscus.
 The trajectory is colour-coded with the instantaneous swimming speed $\bar{v}_d$.
 (c) Variations of the separation distance ($\bar{D}_s$) between the active droplet and the meniscus along the curvilinear coordinate ${l^*_c}$ for different meniscus geometry $(\alpha_c)$.
 (d) (I-II) Temporal variations of $\bar{v}_d$ as the active droplet approaches, swims along, and leaves the meniscus.
 The data is colour-coded with $l^*_c$, the droplet location along the meniscus.
 The grey colour denotes the instants before arriving at and after leaving the meniscus corners.  
 The red dashed lines indicate $\bar{v}_d=0.2$. 
 (e) Residence time $\bar{t}_r$ of the droplet at the corners ($w_1$, and $w_2$) of surfactant solution-air ($a-w; \; \lambda=0$) and surfactant solution-oil ($o-w; \; \lambda=10^4$) menisci for different values of $\alpha_c$.}
\end{figure}

In our experiments, confined menisci of aqueous, supramicellar TTAB (cationic surfactant) solution are created inside polydimethylsiloxane (PDMS) microchannels (Fig.~\ref{fgr:1}(a)) of width $2w \approx 80~\mu$m or $160~\mu$m and height $h \approx 50~\mu$m. 
Isotropic CB15 oil droplets of radius $R_d \approx 20\pm1.8\;\mu$m, solubilizing in the TTAB solution are used as a model system for autophoretic microswimmers \cite{toyota2009_selfpropelled, Swarmingbehavior2011,peddireddy2012solubilization,Swimmingdroplets2016, michelin2023_selfpropulsion} (see Supplementary Material (SM), sections~\ref{sec:EM}-A and \ref{sec:EM}-B for experimental details).
Spontaneous symmetry breaking of the surfactant concentration field during solubilization drives interfacial flows that propel the droplets \cite{michelin2013spontaneous,peddireddy2012solubilization,herminghaus2014interfacial,michelin2023_selfpropulsion}.
Near the channel walls, the mean swimming speed of the self-propelling active droplet is measured as $v_w\approx23\pm2.3\;\mu$m/s. 
Bright-field and fluorescence microscopy are used to quantify swimming trajectories, speeds, orientations, and hydrodynamic signatures of the active droplets (see SM, Secs.~\ref{sec:EM}-C and \ref{sec:EM}-D).
The instantaneous swimming velocity of the active droplet in the quasi-2D setup $(h/2R_d \sim 1.2)$ is defined as $\mathbf{v}_d=v_d\mathbf{\hat{e}}=v_d(\cos \psi\mathbf{\hat{x}}+\sin \psi\mathbf{\hat{y}})$, where $\mathbf{\hat{e}}$ is the instantaneous swimming orientation.
All quantities reported below are non-dimenisonalised using $R_d$, $v_w$, and $R_d/v_w$ as the characteristic length, velocity and time scales, respectively.

We consider two meniscus types: surfactant solution–air interfaces ($\lambda = 0$) and surfactant solution–oil interfaces ($10^4$ cSt silicone oil; $\lambda = 10^4$) (see SM, Sec.~\ref{sec:EM}-B). 
Here, $\lambda$ denotes the viscosity ratio between the outer fluid (Fluid 2) and the surfactant solution (swimming medium; Fluid 1) (Figs.~\ref{fgr:1}(a)).
The geometry of the confined meniscus is characterized by the dimensionless parameter $\alpha_c = w\sqrt{\frac{\kappa}{R_d}}$, where $\kappa = \frac{\cos \theta_c}{w}$ is the meniscus curvature, and $\theta_c$ is the contact angle at the microchannel walls (Figs.~\ref{fgr:1}(a) and (b)).
The parameter $\alpha_c$ increases with decreasing $\theta_c$; for fixed $R_d$, larger $\alpha_c$ therefore corresponds to narrower corners formed by the meniscus (see SM, section \ref{sec:EM}-B, Fig.~\ref{fig:S2}). 
The corners near walls 1 and 2 are referred to as $w_1$ and $w_2$ respectively (Figs.~\ref{fgr:1}(b)).

A typical trajectory of an active droplet near a surfactant solution–air meniscus ($\lambda = 0$) is shown in Fig. ~\ref{fgr:1}(b). 
The droplet swims along wall 1 toward the confined meniscus, turns at the $w_1$ corner, and then proceeds along the meniscus to the $w_2$ corner. 
Thus, by design, our experimental set-up enables probing active droplet dynamics near both meniscus corners, each reached with a different approach angle within a single run. 
Interestingly, we find that the active droplet gets attracted towards both corners during its swimming path.
The effect is more pronounced near $w_2$: rather than following the straight line trajectory (dashed line in Fig.~\ref{fgr:1}(b)) and moving away from the meniscus, the droplet reorients (at $\approx 5 R_d$ away from the $w_2$ corner apex; orange circle) and swims toward the corner (see Video~\ref{sec:Videos}-S1).

To quantify this behaviour, we plot the separation distance $\bar{D}_s$ (Fig.~\ref{fgr:1}(b)) from the meniscus as a function of the curvilinear meniscus coordinate $l_c^*$ in Fig.~\ref{fgr:1}(c).
As the droplet departs from the $w_1$ corner ($l_c^*=0$) and swims along the interface, $\bar{D}_s$ initially increases, followed by a sharp decrease near the $w_2$ corner representing the attraction.
The attraction towards the corner becomes accentuated for narrower corners, i.e., at higher $\alpha_c$, obtained by reducing $\theta_c$ (Fig.~\ref{fgr:1}(c); see SM Fig.~\ref{fig:S5}(a) for additional data).

Attraction is also present at the $w_1$ corner, which is more clearly revealed by the droplet swimming speed $\bar{v}_d$ (Fig.~\ref{fgr:1}(d)).
During its trajectory, $\bar{v}_d$ decreases to near zero twice, corresponding to the approach to the two meniscus corners, indicating a transient trapping prior to eventual escape (see Video~\ref{sec:Videos}-S1; also SM Fig.~\ref{fig:S5}(b) for additional data). 
We define the residence time $\bar{t}_r$ at the corner as the interval over which $\bar{v}_d<0.2$ (red dashed lines in Fig.~\ref{fgr:1}(d)), which also coincides with negligible variation in $l_c^*$ (colourbar in Fig.~\ref{fgr:1}(d)). 
$\bar{t}_r$ at both $w_1$ and $w_2$ corners increases monotonically with increasing $\alpha_c$ (Fig.~\ref{fgr:1}(d) and (e)). 
The capture and subsequent release of active droplets are also observed at the corners of a confined surfactant solution-oil meniscus ($\lambda=10^4$; see Video~\ref{sec:Videos}-S1; SM Fig.~\ref{fig:S6}).
For comparable $\alpha_c$, however, $\bar{t}_r$ at both corners is larger for $\lambda=0$ than for $\lambda=10^4$ (Fig.~\ref{fgr:1}(e)). 

\begin{figure}[h!]
 \centering
 \includegraphics[width=8.6cm]{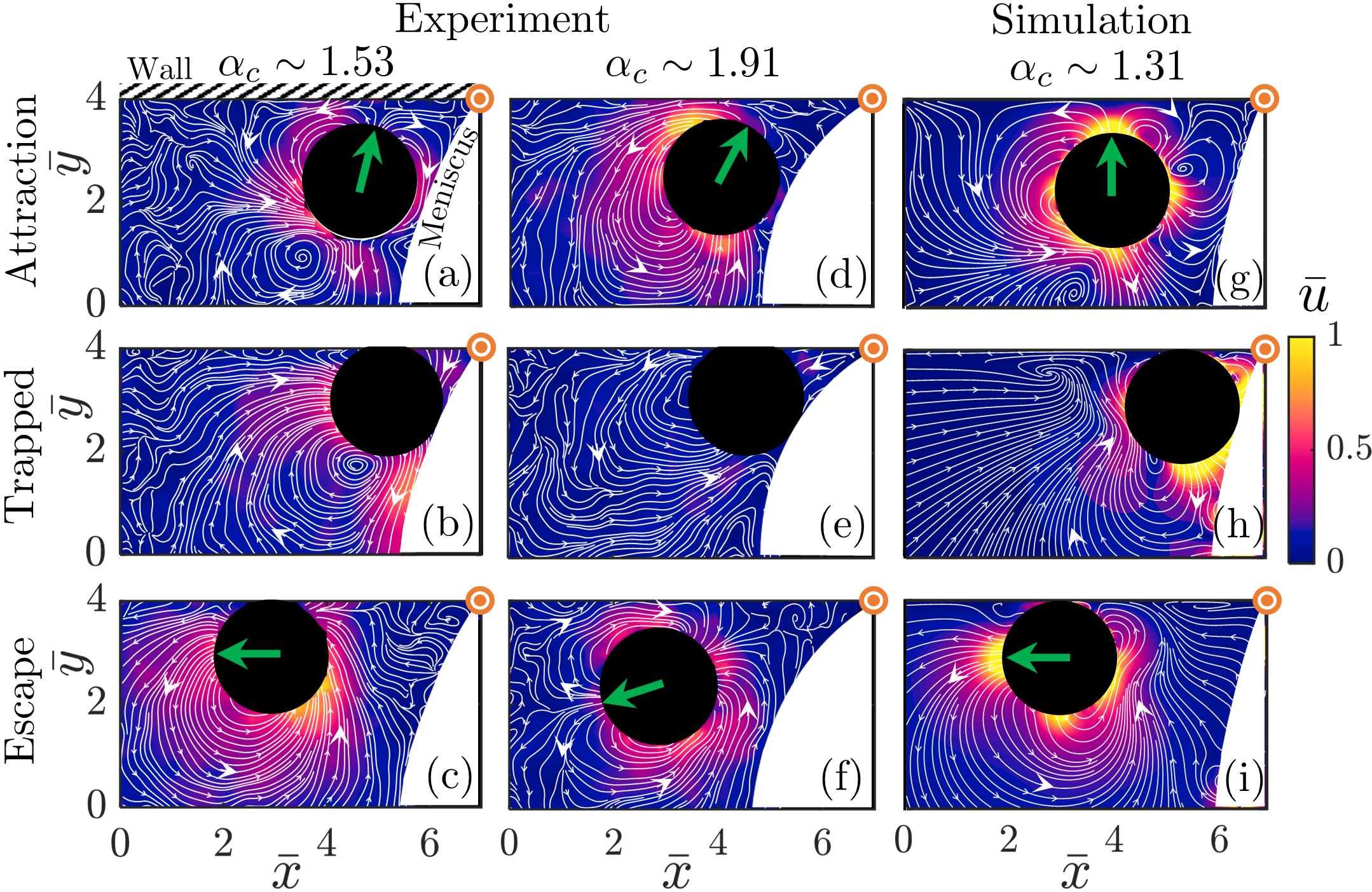}
\caption{\label{fgr:2}Flow fields generated by the active droplet (represented by streamlines; local flow speed $\bar{u}$ as filled contour plots) as it gets attracted to, trapped at, and escapes from the $w_2$ corner of surfactant solution-air meniscus $(\lambda=0)$ for (a)-(c) $\alpha_c = 1.53$ ($\theta_c=55^\circ$), and (d)-(f) $\alpha_c = 1.91$ ($\theta_c=25^\circ$). (g)-(i) Corresponding flow fields obtained from the hybrid lattice Boltzmann Method based numerical simulations of a model microswimmer (squirmer; $\beta=-1$) near a corner ($\alpha_c = 1.31$, $\theta_c=65^\circ$; $\lambda=0.16$) of a confined meniscus.} 
\end{figure}

We next examine how the flow generated by the active droplet—behaving as a moderate pusher far from boundaries—changes as it passes a meniscus corner.
For clarity, we divide the trajectory into three stages: approach, trapping, and escape. 
The evolution of the flow near the $w_2$ corner is shown in Fig.~\ref{fgr:2} (see Video~\ref{sec:Videos}-S2; SM Figs.~\ref{fig:S7} and ~\ref{fig:S8} for the corresponding results at $w_1$ and for $\lambda = 10^4$).
During approach, the flow field is quadrupolar or dipolar depending on $\alpha_c$ (Figs.~\ref{fgr:2}(a) and \ref{fgr:2}(d)).
Larger $\alpha_c$ also corresponds to larger flow speeds $(\bar{u})$ in the droplet's vicinity (compare Fig.~\ref{fgr:2}(a) with \ref{fgr:2}(d)).
When trapped, the stationary droplet exhibits a monopolar flow accompanied by a persistent clockwise (Fig.~\ref{fgr:2}(b)) or anticlockwise (Fig.~\ref{fgr:2}(e)) vortex sustained by surface activity.
Upon escape, the flow recovers a dipolar structure (Figs.~\ref{fgr:2}(c) and \ref{fgr:2}(f)). 
The hydrodynamic signature therefore depends sensitively on the corner geometry and on the droplet's position and orientation within it, and is thus coupled to its dynamics.

To test this interpretation, we simulate a pusher-type squirmer placed at equivalent positions and orientations in a confined meniscus corner using a hybrid lattice Boltzmann–diffuse-interface method (see SM section~\ref{sec:NM} for details).
The simulations reproduce the observed flow profiles (compare Fig.~\ref{fgr:2}(g)–(i) with \ref{fgr:2}(a)–(c)), indicating that hydrodynamics governs the droplet dynamics in this system.
We therefore construct a hydrodynamic theory for microswimmers near meniscus corners that explains the experimental observations and makes additional predictions, as discussed below.

\begin{figure}[h!]
 \centering
 \includegraphics[width=8.3cm]{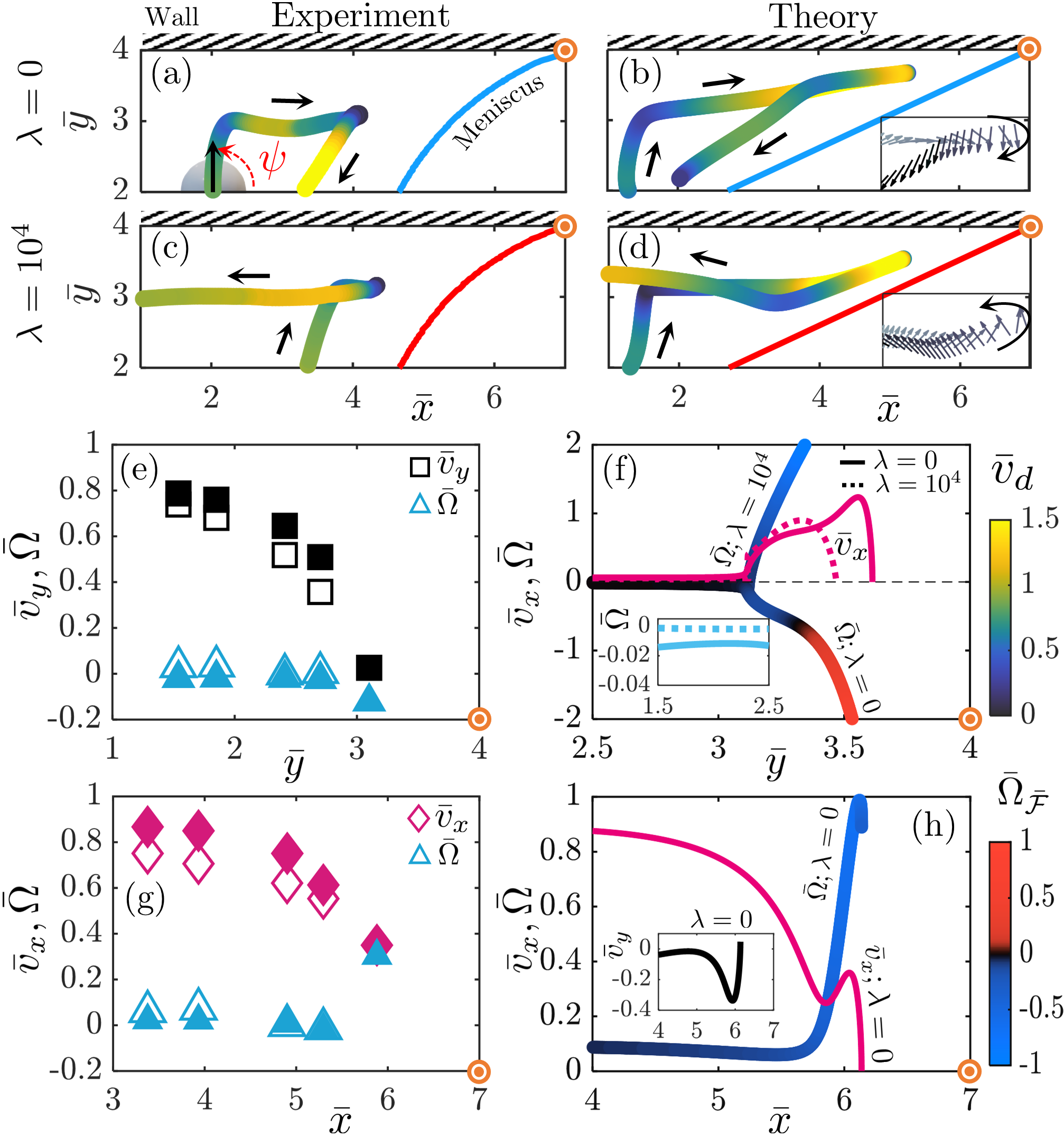}
 \caption{\label{fgr:3}Predictions of the hydrodynamic model. 
Comparison of trajectories between the active droplet from experiments and the microswimmer ($\beta=-1, \bar{\mathcal{S}}=0.4$) from the model near the $w_2$ corner $(\theta_c \approx 25^\circ; \alpha_c=1.90)$  of a confined surfactant solution-air $(\lambda=0)$ ((a), (b)) and oil meniscus $(\lambda=10^4)$ ((c), (d)). 
Arrows in insets show the temporal evolution of the swimming orientation in the corner.
Variations in (e) the transverse velocity component $(\bar{v}_y)$ and angular velocity $(\bar{\Omega})$  with the transverse location $(\bar{y})$ as the microswimmer approaches the $w_2$ corner $(\theta_c= 65^\circ; \alpha_c=1.30)$ for $\lambda=0$, and in (g) the axial velocity $\bar{v}_x$ and $\bar{\Omega}$ with axial position $(\bar{x})$ as the microswimmer approaches the $w_1$ corner; filled and hollow markers are from theory and hybrid lattice Boltzmann simulations respectively. 
(f) Variations in $(\bar{v}_x)$ and $(\bar{\Omega})$ with $\bar{y}$ for different $\lambda$ near the $w_2$ corner;
$\bar{\Omega}$ is colour-coded with the contribution from the force dipole term $(\bar{\Omega}_{\bar{\mathcal{F}}})$. 
The inset shows the initial variation of $\bar{\Omega}$.
(h) Variations of $\bar{v}_x$ and $\bar{\Omega}$ with $\bar{x}$ for $\lambda=0$ near the $w_1$ corner; $\bar{v}_y$ is shown in the inset.}
\end{figure}

In the hydrodynamic model (see SM, section~\ref{sec:TM} for details) we construct the total velocity field near the meniscus corner as
\begin{align}
\label{eqn:velocity}
    \bar{\mathbf{u}}= \bar{\mathbf{u}}_S&+\bar{\mathcal{F}}\bar{\mathbf{u}}_{fd,W}^* + \bar{\mathcal{S}} \bar{\mathbf{u}}_{sd,W}^*\nonumber\\
    &+\bar{\mathcal{F}}\bar{\mathbf{u}}_{fd,I}^*(\theta_c;\lambda) + \bar{\mathcal{S}} \bar{\mathbf{u}}_{sd,I}^*(\theta_c; \lambda)
\end{align}
The active droplet is modelled as a pusher-type microswimmer whose velocity field is approximated by the superposition of an axisymmetric force dipole (representing the force-free swimming) and a source dipole (accounting for the finite size of the microswimmer) flow singularities \cite{JFM_Lauga_2012, POF_Lauga_2014, KimKarrila}-- $\bar{\mathbf{u}}_S=\bar{\mathcal{F}} \bar{\mathbf{u}}_{fd} + \bar{\mathcal{S}} \bar{\mathbf{u}}_{sd}$, where $\bar{\mathcal{F}}$ and $\bar{\mathcal{S}}$ are the strengths of the force and source dipoles respectively; $\bar{\mathcal{F}}\propto- \beta \bar{\mathcal{S}}$, where $\beta$ is the swimmer parameter ($\beta<0$ for pushers and  $\beta>0$ for pullers). 
The influence of the confined meniscus corner is incorporated by superimposing the hydrodynamic image systems \cite{Blake_1971, Blake1974, JFM_Lauga_2012, POF_Lauga_2014} for each singularity corresponding to a horizontal, `no-slip' wall producing $(\bar{\mathbf{u}}_{fd, W}^*; \bar{\mathbf{u}}_{sd, W}^*)$ and a fluid-fluid interface inclined to the wall at an angle $\theta_c$ producing $(\bar{\mathbf{u}}_{fd, I}^*(\theta_c; \lambda); \bar{\mathbf{u}}_{sd, I}^*(\theta_c; \lambda))$.

The components $(\bar{v}_x,\bar{v}_y)$ of the resulting translational velocity of the microswimmer are obtained using Fax\'en's first law \cite{JFM_Lauga_2012, POF_Lauga_2014, Byjesh_PRF, KimKarrila}.
See SM Section~\ref{sec:TM} for the final equations.
The angular velocity is obtained from Fax\'en's second law for a torque-free microswimmer \cite{JFM_Lauga_2012, POF_Lauga_2014, KimKarrila, Byjesh_PRF} as 
\begin{align}
\label{eqn:omega}
\bar{\Omega}=\mp\bar{v}_o \bar{\mathcal{F}}&\left[\frac{3\sin2\psi}{16\bar{y}^3}+\frac{3\sin{2(\theta_c+\psi)}}{16\bar{l}^3}\right]\notag\\
&\quad
\pm\bar{v}_o\bar{\mathcal{S}}\left[\frac{3\cos\psi}{8\bar{y}^4}-\frac{3\lambda\cos{(\theta_c+\psi)}}{8(1+\lambda)\bar{l}^4}\right]
\end{align}
where $\pm/\mp$ corresponds to the $w_1$(top symbol) and $w_2$ (bottom symbol) corners; and $\psi$ is the swimmer orientation. 
Terms involving $\bar{y}$ (the shortest distance from the wall) and $\bar{l}$ (the shortest distance from the meniscus, $D_s+R_d$) represent the hydrodynamic interactions of the microswimmer with the wall and the meniscus, respectively.

For a moderate, pusher-type microswimmer $(\beta=-1)$, the predicted trajectories near the $w_2$ corner (see Video~\ref{sec:Videos}-S3) agree well with those from experiments for both $\lambda=0$ (Figs.~\ref{fgr:3}(a) and (b)) and $\lambda=10^4$ (Figs.~\ref{fgr:3}(c) and (d)).
While both cases exhibit attraction and eventual escape, the underlying dynamics differ.

For $\lambda=0$, a distant microswimmer initially oriented at $\psi_i = 90^\circ$ undergoes clockwise rotation ($\bar{\Omega}<0$) towards the corner due to the hydrodynamic interactions of the pusher-type force dipole with the surfactant solution-air interface (second term in Eq.~\ref{eqn:omega}; Fig.~\ref{fgr:3}(e), and solid blue line in inset in Fig.~\ref{fgr:3}(f)).
The finite size (source dipole) of the microswimmer plays no role here because its image $\bar{\mathbf{u}}_{sd,I}^*(\theta_c;\lambda=0)$ is also a source dipole that produces an irrotational flow.
Further, during this stage of approach (till $\bar{y} \sim 3$), $\bar{v}_d$ (dominated by $\bar{v}_y$) gradually decreases due to hydrodynamic interactions with the wall (Fig.~\ref{fgr:3}(e)), and explains the gradual reduction in $\bar{v}_d$ of the active droplet as it approaches $w_2$ corner (Fig.~\ref{fgr:1}(d)).

Near the corner, the finite size of the microswimmer becomes important.
Hydrodynamic interactions between the source dipole and the wall induce additional clockwise rotation (third term in Eq.~\ref{eqn:omega}; Fig.~\ref{fgr:3}(f)), and generate a positive axial component of velocity $\bar{v}_x$ (solid red line in Fig.~\ref{fgr:3}(f)) which further reorients and attracts the microswimmer into the corner. 
Eventually, the force dipole induced hydrodynamic interactions sharply reduce $\bar{v}_x$ (Fig.~\ref{fgr:3}(f)), and hence $\bar{v}_d$ (as also seen in the immediate vicinity of the corner in Fig.~\ref{fgr:1}(d)), leading to trapping of the microswimmer.
However, the source dipole induced $\bar{\Omega}$ continues to reorient the microswimmer clockwise, interestingly, against the anticlockwise rotation arising from the force dipole interactions ($\bar{\Omega}_{\bar{\mathcal{F}}}>0$; colorbar in  Fig. \ref{fgr:3}(f)), enabling the microswimmer to escape from the $w_2$ corner (note arrows in the inset of Fig. \ref{fgr:3}(b); also see Video~\ref{sec:Videos}-S3).

For $\lambda=10^4$, the image systems differ, modifying the dynamics.
In this case, even far from the corner, the finite size of the microswimmer introduces an additional anticlockwise contribution ($\bar\Omega>0$; fourth term in Eq.~\ref{eqn:omega}) due to hydrodynamic interactions with the surfactant solution-oil interface.
This reduces the net $\bar{\Omega}$ (dotted blue line in the inset in Fig.~\ref{fgr:3}(f)) leading to a weak (slow) attraction towards the corner.
Here, close to the corner, source dipole interactions with both the wall and the interface generate an anticlockwise rotation ($\bar{\Omega}>0$; Fig.~\ref{fgr:3}(f)) and a positive $\bar{v}_x$ (dotted red line in Fig.~\ref{fgr:3}(f)) that attract the microswimmer into the corner.
Continued positive $\bar{\Omega}$ reorients the microswimmer against the trapping effect of the force dipole and interface interaction (reduction in $\bar{v}_x$, and $\bar{\Omega}_{\bar{\mathcal{F}}}<0$ in this case; Fig.~\ref{fgr:3}(f)).
Accordingly, the microswimmer eventually escapes from the corner following an anticlockwise change in orientation (note arrows in the inset of Fig.~\ref{fgr:3}(d); also see Video~\ref{sec:Videos}-S3).

The larger $|\bar{\Omega}|$ near the corner for $\lambda=10^4$, (compare plots in Fig.~\ref{fgr:3}(f)) results in faster escape of the microswimmer, consistent with the smaller $\bar{t}_r$ observed experimentally (Fig.~\ref{fgr:1}(e); also see SM, Fig.~\ref{fig:S10}(a)).
Larger $|\bar{\Omega}|$ with decreasing $\alpha_c$ (see SM, Fig.~\ref{fig:S10}(b)), due to weakened force dipole interactions in a wider corner geometry, is also consistent with decreasing $\bar{t}_r$ observed in the experiments (Fig.~\ref{fgr:1}(e)).  

For a microswimmer approaching the $w_1$ corner along the wall $(\psi_i=0)$, the dynamics are similar for both $\lambda=0$ and $\lambda=10^4$ (for comparison with experiments, see SM, Fig.~\ref{fig:S9}), unlike the $w_2$ corner.
In the proximity of $w_1$, hydrodynamic interactions of the force dipole and source dipole with the interface significantly reduce $\bar{v}_x$ (Fig.~\ref{fgr:3}(g), (h)).
Simultaneously, $\bar{v}_y$ becomes negative (inset in Fig.~\ref{fgr:3}(h)) mainly due to source dipole induced interactions in the meniscus corner. This directs the microswimmer into the corner, with an overall reduction in $\bar{v}_d$ (as seen in Fig.~\ref{fgr:1}(d)).
Eventually, $\bar{v}_x$ and $\bar{v}_y$ become zero (Fig.~\ref{fgr:3}(h)) trapping the microswimmer at the corner.
However, as in $w_2$, the hydrodynamic interaction induced by the source dipole with the wall (for $\lambda=0$) and also with the meniscus (for $\lambda=10^4$) reorients the microswimmer away from the corner by generating a high positive $\bar{\Omega}$ (Fig.~\ref{fgr:3}(g), (h)) allowing it to escape.

The robustness of the model and its quantitative predictions for  $\bar{v}_x$, $\bar{v}_y$ and $\bar{\Omega}$ near the meniscus corners are confirmed by performing analogous hybrid lattice Boltzmann numerical simulations (empty markers in Fig.~\ref{fgr:3}(e),(g)).

\begin{figure}[h!]
 \centering
 \includegraphics[width=8.6cm]{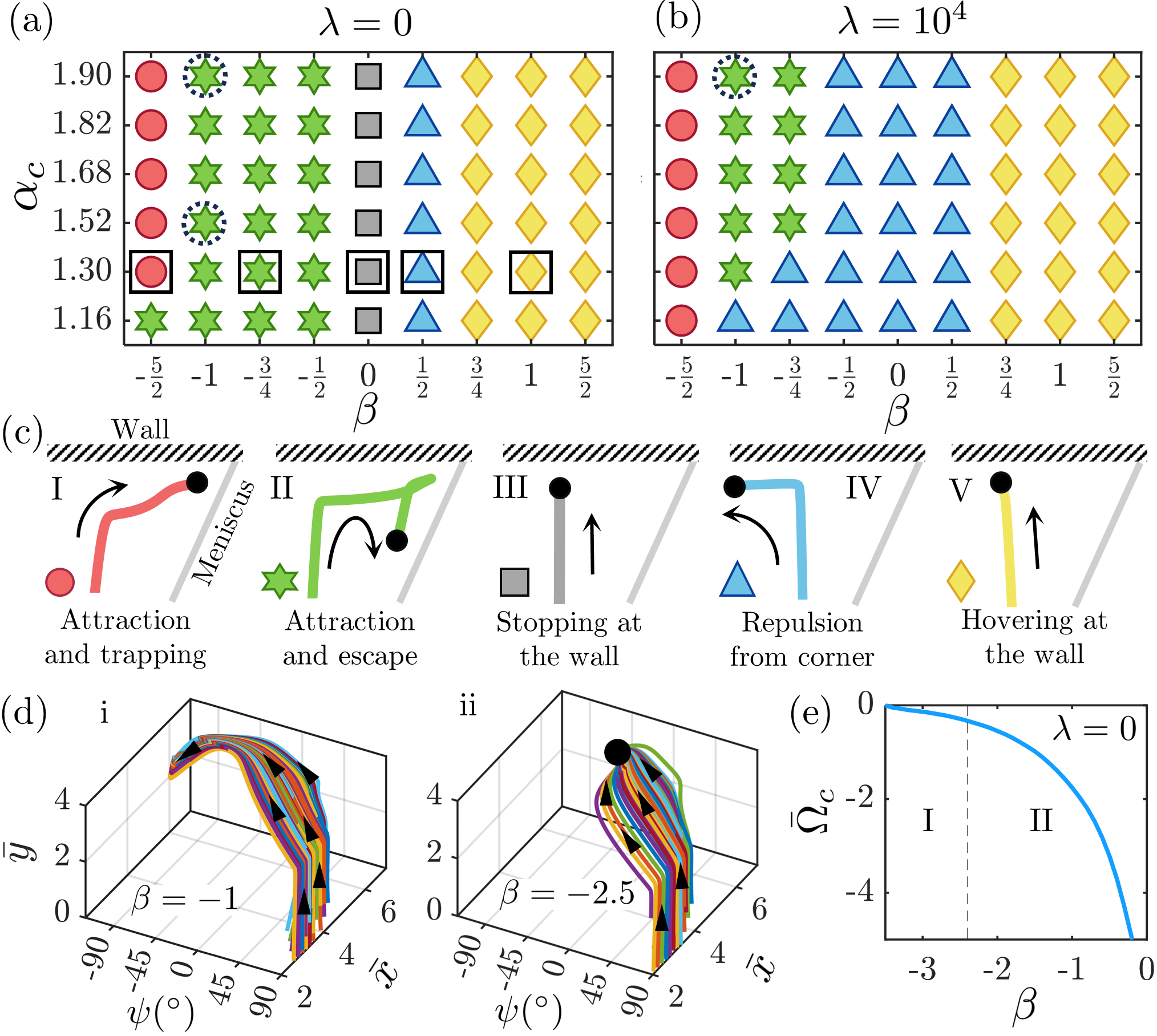}
 \caption{\label{fgr:4} State diagrams for the trajectory types at $w_2$ corner for (a) $\lambda=0$, and (b) $\lambda=10^4$.
 The symbols correspond to the 5 different trajectory types as explained in (c).  
(d) Ensemble of microswimmer trajectories near the $w_2$ corner for $\lambda=0$ and $\alpha_c=1.30$ in the $x-y-\psi$ phase space covering a range of initial conditions for (i) $\beta=-1$, and (ii) $\beta=-2.5$. 
(e) Variation of the non-dimensional angular velocity $(\bar{\Omega}_c)$ at the nearest point to the apex of the meniscus corner $(\alpha_c=1.30)$ with $\beta$.
I and II mark the existence of the respective trajectory types as in (c).} 
\end{figure}

Motivated by the agreement between the hydrodynamic model, experiments and simulations, we use the model to make further predictions. 
The analysis focuses on the $w_2$ corner over the parameter space $(\beta, \alpha_c)$; similar discussions for the $w_1$ corner are in SM (Figs.~\ref{fig:S13} and ~\ref{fig:S14}).
Five distinct trajectory types (Fig.~\ref{fgr:4}(c)) arise near the corner: attraction followed by trapping (type I) or escape (type II), no attraction with arrest at the wall (type III), repulsion from corner (type IV), and hovering near the wall (type V). 
The state diagrams Fig.~\ref{fgr:4}(a) and~\ref{fgr:4}(b) map the occurrence of these trajectories for $\lambda=0$ and $\lambda=10^4$, respectively.

For $\lambda = 0$ (Fig.~\ref{fgr:4}(a)), pushers $(\beta<0)$ are attracted to the corner, whereas pullers $(\beta>0)$ are not; the latter is also true for $\lambda = 10^4$ (Fig.~\ref{fgr:4}(b)).
A neutral microswimmer $(\beta=0)$ does not exhibit attraction, and it instead stops at the wall (type III), indicating the necessity of far field pusher-type force dipole interaction for attraction.
For moderate to large $\alpha_c$, the corner-attracted pushers exhibit either `escape' (type II) or `trapped' (type I).
The transition between types II and I occurs with increasing force dipole strength ($\beta<-2.4$), and arises from the dependence of $\bar{\Omega}$ on $\beta$ (Fig. \ref{fgr:4}(e)).
When force and source dipole contributions become comparable, they negate each other, and $\bar{\Omega}$ is reduced, leading to trapping.
However, for smaller $\alpha_c$ $(\leq 1.16)$, even strong pushers exhibit `attraction and escape' (type II) due to larger $|\bar{\Omega}|$ arising from weakened force dipole interactions in the wider corner geometry.

For $\lambda=10^4$ (Fig.~\ref{fgr:4}(b)), the same trajectory types are observed, but their occurrence and underlying mechanisms differ. 
Along with neutral swimmers, weak pushers are now also repelled away from the corner (type IV) due to the additional source dipole-induced rotation away from the interface. 
As for $\lambda=0$, for moderate to large $\alpha_c$, pushers exhibit either attraction followed by escape (type II) or trapping (type I), depending on the force dipole strength.
However, for $\alpha_c\leq 1.16$, the escape regime is absent, and the pusher-type swimmers directly transition to trapping due to reduced $|\bar{\Omega}|$ near the corner.
Taken together, these results show that, with the gradual increase in the viscosity of the outer fluid, repulsion from the corner is enhanced: first weak pushers behave like neutral swimmers, followed by even moderate pushers (see SM, Fig.~\ref{fig:S11}).
Our analysis shows that the state diagram in ($\beta, \alpha_c$) becomes effectively invariant to $\lambda$ beyond $\lambda \sim \mathcal{O}(10)$ (compare SM, Fig.~\ref{fig:S11}(c) and Fig.~\ref{fgr:4}(b)).
Consequently, the behaviour at $\lambda=10^4$ approximates that near a corner formed by two solid walls (compare SM, Fig.~\ref{fig:S11}(d) and Fig. \ref{fgr:4}(b); also see SM, Fig.~\ref{fig:S12}).   

Finally, Fig.~\ref{fgr:4}(d) shows an ensemble of trajectories in the $x-y-\psi$ phase space, demonstrating the robustness of the observed change in swimming behaviour across different initial conditions.
For a moderate pusher $(\beta=-1)$, trajectories exhibit attraction and escape (type II) over a wide range of initial conditions (Fig.~\ref{fgr:4}(d)(i)). 
For a strong pusher $(\beta=-2.5)$, all trajectories converge to a fixed point (black circle in Fig.~\ref{fgr:4}(d)(ii)) corresponding to trapping (type I). 
The transition from escape to trapping thus corresponds to the emergence of a stable fixed point in the phase space, driven by the reduction in $\bar{\Omega}$ with increasing $|\beta|$ (Fig. \ref{fgr:4}(e)). 

In conclusion, we have demonstrated that pusher-type microswimmers, like self-propelling active droplets, are attracted towards meniscus corners in microconfinements, followed by transient trapping and escape.
Our analysis shows that hydrodynamic interactions between the pusher-type force dipole and the wall-interface corner attract the microswimmer and favour trapping; however, the finite size (source dipole)-induced hydrodynamic interactions mainly promote escape by countering the force dipole effects.
Based on such hydrodynamic interactions, we show the existence of a rich zoology of microswimmer trajectories near the meniscus corner which can be accessed by tuning the type and strength of the microswimmer, corner geometry,  and the viscosity ratio for the liquid interface.
Our work provides a general framework for understanding the dynamical features of low-Reynolds-number swimming near corner geometries spanning the spectrum of wall-interface to wall-wall configurations of different opening angles and swimmer types.  
Finally, the results offer simple guidelines for  sorting microswimmers, both biological and synthetic, depending on their hydrodynamic signature by tuning corner geometries within microfluidic setups.  

\begin{acknowledgments}
\textit{Acknowledgements}--S.G. acknowledges the Prime Minister's Research Fellowship (PMRF), a scheme by the Government of India to improve the quality of research in various research institutions in the country.
R.D. acknowledges support from Science and Engineering Research Board (SERB) (now subsumed under Anusandhan National Research Foundation), Department of Science and Technology (DST), Government of India, through Grant No. SRG/2021/000892, and from Indian Council of Medical Research (ICMR) through Grant No. IIRP-2023-2832. 
\end{acknowledgments}
\bibliography{corner}

\newpage
\clearpage
\widetext
\renewcommand{\thefigure}{S\arabic{figure}}
\renewcommand{\theHfigure}{S\arabic{figure}}  
\setcounter{figure}{0}
\renewcommand{\thesection}{S\Roman{section}}
\setcounter{section}{0}

\begin{center}
   
    {\large \bfseries Supplementary material for Hydrodynamic capture and release of a microswimmer by a meniscus corner}\\[2ex]
   
\end{center}
\tableofcontents

\clearpage
\twocolumngrid
\section{EXPERIMENTAL METHODOLOGY}
\label{sec:EM}
\subsection{Active droplet generation}
\begin{figure}[h!]
  \centerline{\includegraphics[width=8.6cm]{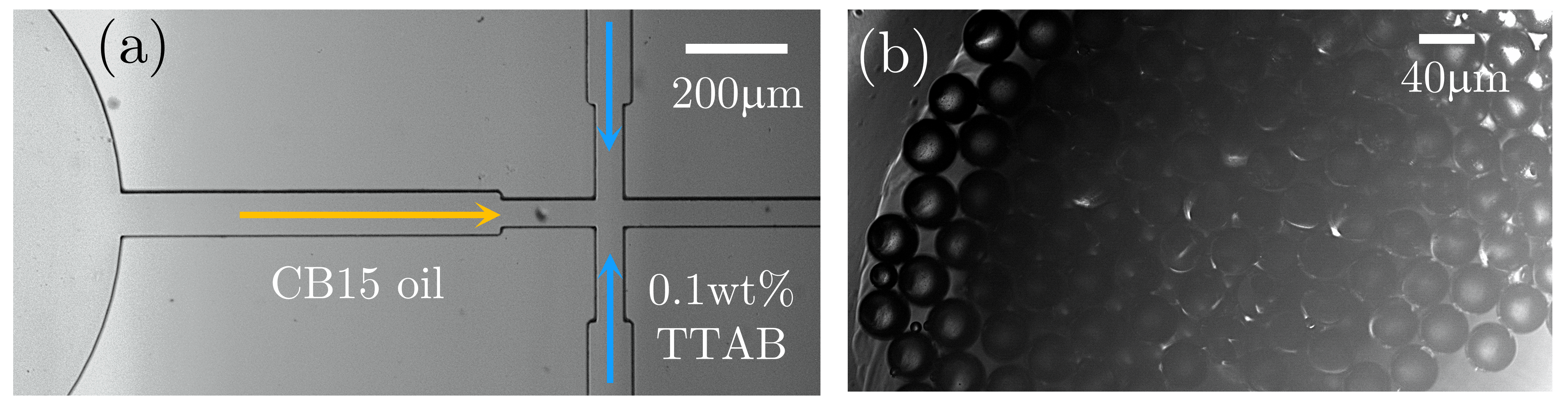}}
  \caption{(a) The bright-field image of the droplet generator made of 10:1 PDMS. (b) The bright-field image of monodisperse active CB15 oil droplets.}
\label{fig:S1}
\end{figure}

We use a flow-focusing T-junction shaped device (see Fig.~\ref{fig:S1}(a)) to generate (S)-4-Cyano-$4'$-(2-methylbutyl) biphenyl (CB15, PureSynth) oil droplets. As shown in Fig.~\ref{fig:S1}(a), from one end we pump CB15 oil with a flow rate of $\approx1\;\mu$l/hr, and from the other two ends, we give a flow rate of $\approx200\;\mu$l/hr to an aqueous solution of tetradecyltrimethyl ammonium bromide ($0.1\;\text{wt}\%$ TTAB, Sigma Aldrich). One can tune the flow rate to get the respective droplet radius. After the formation of monodisperse (Fig.~\ref{fig:S1}(b)) droplets, we store those in a container with $0.1\; \text{wt} \% $ (below critical micelle concentration (CMC) =
0.13 wt$\%$) TTAB solution.

\subsection{Meniscus formation}
\begin{figure}[h!]
  \centerline{\includegraphics[width=8cm]{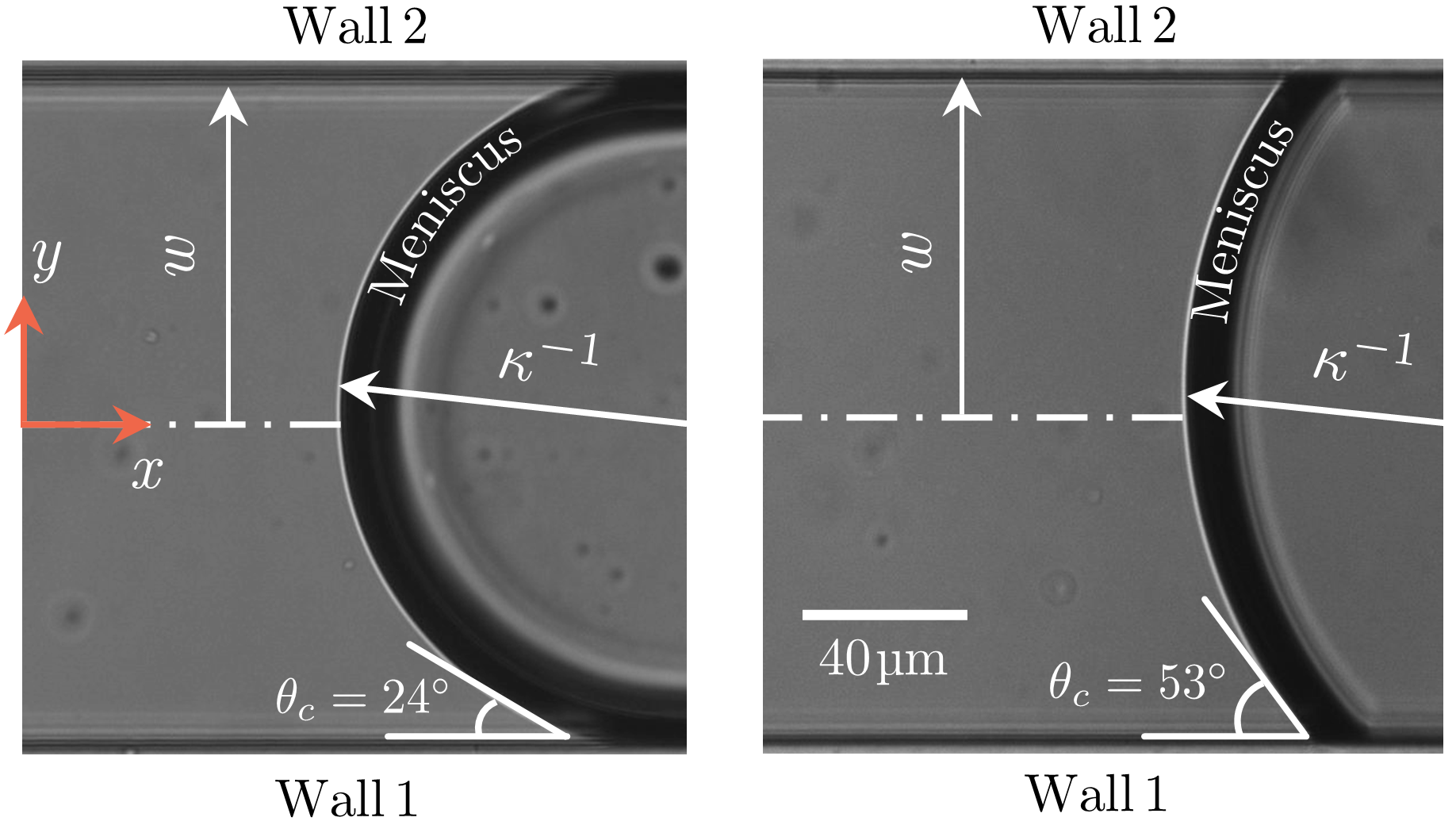}}
  \caption{The bright-field images of two different air-water interfaces having different contact angles inside micro-confinement.}
\label{fig:S2}
\end{figure}

\begin{figure}[H]
\centerline{\includegraphics[width=7cm]{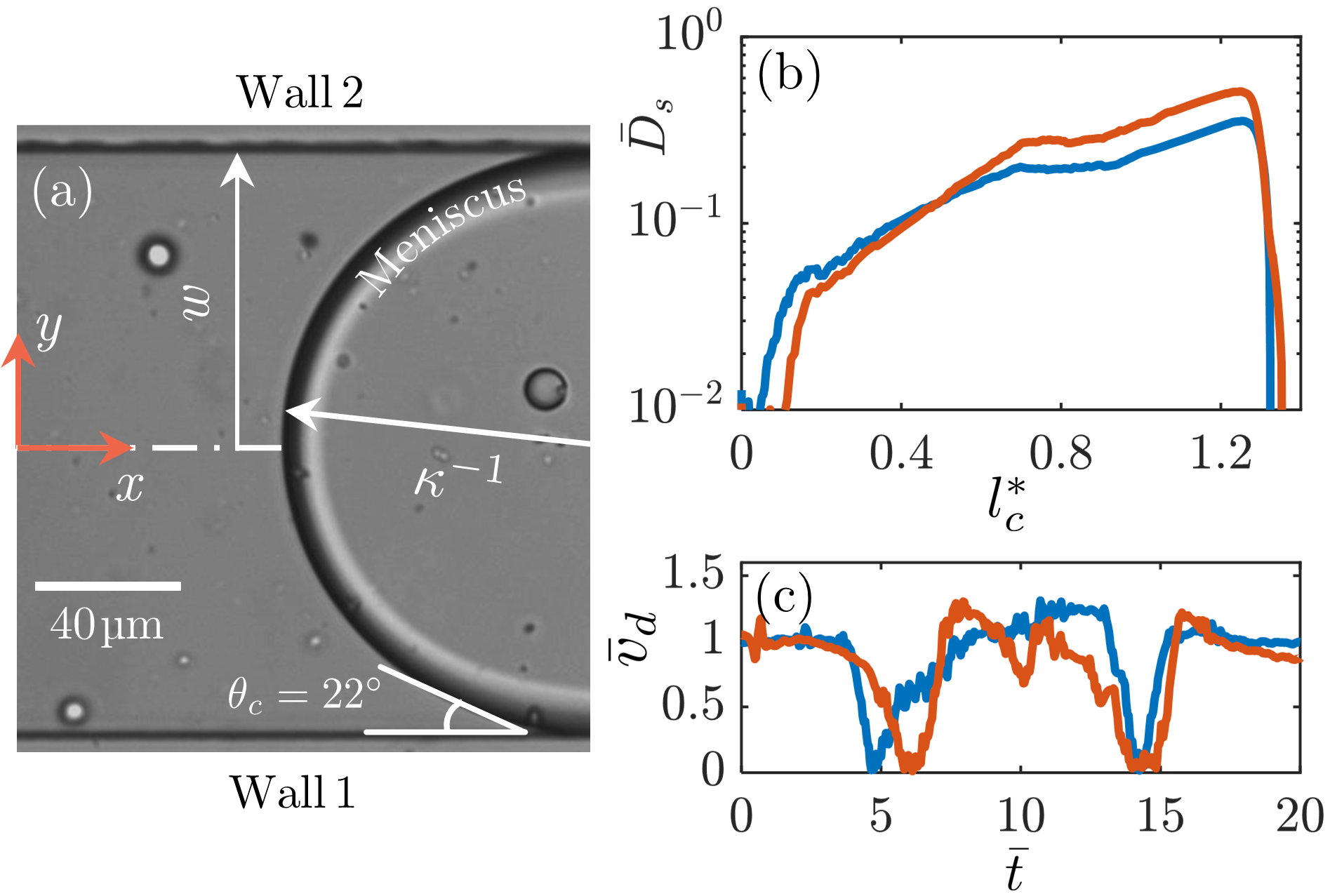}}
  \caption{(a) The bright-field image of surfactant solution-oil interfaces ($\lambda=10^4$) inside microconfinement. (b) Variations of $\bar{D}_s$ along ${l^*_c}$ for multiple experimental outcomes.
 (c) Temporal variations of $\bar{v}_d$ as the microswimmer approaches, swims along, and leaves the oil-water meniscus. Here $\alpha_c\sim 1.9$.}
\label{fig:S6}
\end{figure}

We use $10:1$ PDMS (polydimethylsiloxane mixture of Sylgard 184 prepolymer and cross-linker in the weight ratio of $10: 1$) microchannels (see Fig.~\ref{fgr:1}(a)) of widths $80$ and $160\; \mu$m (depth $\sim50\;\mu$m) to create the air-water meniscus. Simply with a hand pipette, we inject $7.5\;\text{wt} \% $ TTAB solution with CB15 droplet solution (ratio of $2:1$), where the mixed solution volume is approximately $50\%$ of the total volume of the microchannel setup to get a static air-water interface (see Figs.~\ref{fgr:1}(a) and (b) and Fig.~\ref{fig:S2}). We use two different waiting times after plasma oxidation (Plasma Cleaner; Harrick Plasma) (bonding the microchannel with the glass slide). For roughly $30-60$ minutes of waiting time, we get the contact angles ($\theta_c$) around $20^\circ-25^\circ$ (see Fig.~\ref{fig:S2} left), and for 2 hours$+$, $\theta_c\approx 50^\circ-60^\circ$ (see Fig.~\ref{fig:S2} right). \\
For the oil-water case, we first inject the droplet solution into the microchannel; then, with the syringe pump (CHEMYX), we introduce silicon oil ($10^4$ cSt) with a flow rate of $5\;\mu$l/hr. Once the oil reaches the center of the microchannel, we stop the pump and wait until the interface becomes static (see Fig.~\ref{fig:S6}(a)). In this case, $\theta_c$ is always in the range of around $20^\circ-25^\circ$.

\subsection{Bright-field microscopy and droplet's centroid tracking}
By using the bright-field microscopy and a CMOS monochrome digital camera (image size $1920$ pixel $\times 1200$ pixel), we capture the video of the active droplets near the interface at $25$ fps (frame per second, time step $0.04$ sec) with a resolution of $0.176\;\mu$m/pixel ($30\times$ objective). With an in-house MATLAB code, we extract the video into raw images. Next, we convert those into binary images so that the other domains appear black, except for the droplet, interface, and wall. To identify the droplet, we draw a circle around it in the first image and store the area of the drawn region. Subsequently, the MATLAB code automatically calculates the droplet centroid in each frame from the stored area value. 

\subsection{Fluorescence microscopy and Particle Image Velocimetry}
We use high-resolution fluorescence microscopy and the micro Particle Image Velocimetry ($\mu$PIV) technique (visit \cite{PRF_Subhasish_2025} for more details) to resolve the velocity field generated by the active droplets near the meniscus. We use $40\times$ objective (having a resolution of $0.14\;\mu$m/pixel) to capture the video of the flow field (generated by the $200$ nm fluorescence tracer particle (Thermo Fisher Scientific). We explore the flow field using PIVLab software (Interrogation area (pixel): Pass 1: 96, step size 48, and Pass 2: 48, step size 24) and in-house MATLAB code.

\section{THEORETICAL METHODOLOGY}
\label{sec:TM}
\subsection{Modelling of droplet microswimmer as a squirmer}
In this study, we characterise our droplet microswimmer using the squirmer model \cite{blake1971spherical, squirmer2} by adding a force dipole ($fd$) and a source dipole ($sd$), and nondimensionalise its velocity components relative to its near-wall average speed $v_w$. Since the net hydrodynamic force and torque acting on the squirmer are zero, the leading-order term is the force dipole. So we define the velocity profile of the squirmer as follows:
\begin{align}
\label{eqn:1}
\bar{\textbf{u}}_{S}=\mathcal{\bar{F}}\bar{\textbf{u}}_{fd}+\mathcal{\bar{S}}\bar{\textbf{u}}_{sd}
\end{align}
Here, $\mathcal{\bar{F}}$ and $\mathcal{\bar{S}}$ are non-dimensional force dipole and source dipole strengths \cite{strength} which are related to the second and first squirming modes $B_2$ and $B_1$ respectively \cite{squirmer2}.So 
\begin{align}
\label{eqn:2}
\mathcal{\bar{S}}=\frac{\frac{8}{3}\pi\eta_1B_1{R}^3_d}{8\pi\eta_1v_d{R}^3_d}=\frac{B_1}{3v_d}
\end{align}
\begin{align}
\label{eqn:3}\mathcal{\bar{F}}=-\frac{4\pi\eta_1B_2{R}^2_d}{8\pi\eta_1v_d{R}^2_d}=-\frac{B_2}{2v_d}=-1.5\beta\mathcal{\bar{S}}
\end{align}
$\beta(=B_2/B_1)$ represents the squirmer parameter and $\eta_1$ is the viscosity of the swimming medium and $v_d$ denotes the squirmer's instantaneous swimming speed.\\\;\
\textbf{Note:} The outcomes from the theoretical model and the numerical simulation match quite well for $\bar{\mathcal{S}}=0.4$, so we consider $\bar{\mathcal{S}}=0.4$ for our model calculation.
\subsection{Introducing the hydrodynamic interactions}
We use the method of images \cite{JFM_Lauga_2012, POF_Lauga_2014} to study the hydrodynamic interaction (HI) between the squirmer and a corner formed by a wall and an interface. We simply add the velocity field by HI with equation~\ref{eqn:1} to get the total velocity field near a meniscus corner as:
\begin{align}
\label{eqn:4}
\bar{\textbf{u}}&= \bar{\textbf{u}}_{S}+\bar{\textbf{u}}^*_{\text{HI}}\notag\\
\text{with}\; \bar{\textbf{u}}^*_{\text{HI}}&=
\bar{\mathcal{F}}\bar{\mathbf{u}}_{fd,W}^* + \bar{\mathcal{S}} \bar{\mathbf{u}}_{sd,W}^*\nonumber\\
    &+\bar{\mathcal{F}}\bar{\mathbf{u}}_{fd,I}^*(\theta_c;\lambda) + \bar{\mathcal{S}} \bar{\mathbf{u}}_{sd,I}^*(\theta_c; \lambda)
\end{align}
 $\bar{\textbf{u}}_{fd}^*$ and $\bar{\textbf{u}}_{sd}^*$ represent the velocity fields due to interactions with the wall ($W$) and the interface ($I$). First, we solve the image system for a solid wall ($\lambda\rightarrow\infty$) in the $x-y$ plane. Next, for an interface with viscosity ratio $\lambda$ ($=\eta_2/\eta_1$), we solve it along the interface by considering a new plane $x'-y'$ (see Fig.~\ref{fig:S3}) which depends on $\theta_c$ (contact angle). Later, we transform the coordinate to the standard $x-y$ plane. Finally, we add the two terms to obtain the hydrodynamic model for the corner equation~\ref{eqn:4}.
\\\;\
\textbf{Note:} To evaluate the equations of motion, we solve the image system for the centroid location of the squirmer, where the $fd$ and $sd$ velocities (eqn.~\ref{eqn:1}) have a singularity, but this is nothing but the instantaneous swimming velocity of the squirmer $=v_d\hat{\textbf{e}}$. So
\begin{align}
\label{eqn:5}
\textbf{u}_{S}=\textbf{v}_{d}={v}_d(\cos\psi\hat{\textbf{x}}+\sin\psi\hat{\textbf{y}})
\end{align}
\subsection{Swimmer dynamics at meniscus corner}
\begin{figure}[h!]
  \centerline{\includegraphics[width=8cm]{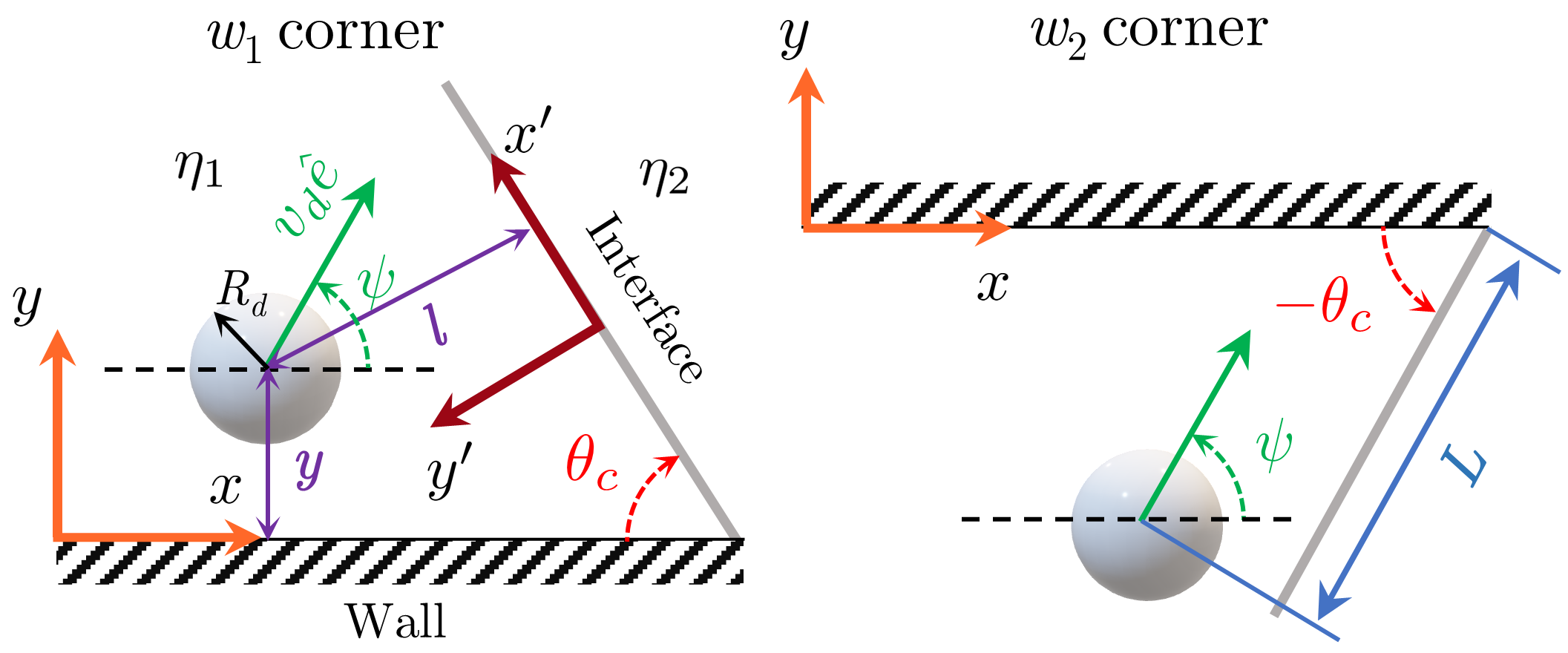}}
  \caption{Schematic of a spherical squirmer approaching the wall-interface corners (both $w_1$ and $w_2$ corners) with a velocity $v_d$ and orientation $\psi$. Here $\eta_1$, $\eta_2$ are the viscosities of the swimming medium and the outer fluid. }
\label{fig:S3}
\end{figure}

We use Faxen's first law \cite{KimKarrila,JFM_Lauga_2012} to resolve the translational velocity components as follows:

\begin{align}
\label{eqn:6}
\bar{v}_{x}&=\bar{v}_o(\cos\psi\pm\bar{\mathcal{F}}\left[\frac{3}{8\bar{y}^2}-\frac{7}{64\bar{y}^4}\right]\sin2\psi\notag\\
&\quad\mp\bar{\mathcal{F}}\left[FD_1\cos{\theta_c}+(FD_2+FD_3)\sin{\theta_c}\right]\\
&\quad\mp\frac{\bar{\mathcal{S}}}{4}\left[\frac{1}{\bar{y}^3}-\frac{3}{4\bar{y}^5}\right]\cos\psi\mp\bar{\mathcal{S}}\left[SD_1\cos{\theta_c}
+SD_2\sin{\theta_c}\right])\notag
\end{align}
\begin{align}
\label{eqn:7}
\bar{v}_{y}&=\bar{v}_o(\sin\psi\pm\bar{\mathcal{F}}\left[\frac{-3+9\sin^2\psi}{8\bar{y}^2}+\frac{7-11\sin^2\psi}{64\bar{y}^4}\right]\notag\\
&\quad\pm\bar{\mathcal{F}}\left[FD_1\sin{\theta_c}-(FD_2+FD_3)\cos{\theta_c}\right]\\
&\quad\mp\bar{\mathcal{S}}\left[\frac{1}{\bar{y}^3}-\frac{1}{8\bar{y}^5}\right]\sin\psi\pm\bar{\mathcal{S}}\left[SD_1\sin{\theta_c}-SD_2\cos{\theta_c}\right])\notag
\end{align}
Where
\begin{align}
\label{eqn:8}
FD_1&=\left[\frac{3\lambda}{8(1+\lambda)\bar{l}^2}-\frac{4+7\lambda}{64(1+\lambda)\bar{l}^4}\right]\sin{2(\theta_c+\psi)},\notag\\
FD_2&=\left[\frac{(2+3\lambda)-3(2+3\lambda)\cos{2(\theta_c+\psi)}}{16(1+\lambda)\bar{l}^2}\right],\notag\\
FD_3&=\left[\frac{(2+3\lambda)+(2+11\lambda)\cos{2(\theta_c+\psi)}}{128(1+\lambda)\bar{l}^4}\right],\\
SD_1&=\left[\frac{1+2\lambda}{8(1+\lambda)\bar{l}^3}+\frac{1-12\lambda}{64(1+\lambda)\bar{l}^5}\right]\cos{(\theta_c+\psi)} \;\text{and}\notag\\
SD_2&=\left[\frac{1+4\lambda}{4(1+\lambda)\bar{l}^3}+\frac{1-2\lambda}{16(1+\lambda)\bar{l}^5}\right]\sin{(\theta_c+\psi)}\notag
\end{align}
Finally, using Faxen's second law \cite{KimKarrila, JFM_Lauga_2012}, we equate the equations for the rotational velocity component:
\begin{align}
\label{eqn:9}
\bar{\mathbf{\Omega}}=\frac{1}{2}\nabla\times\bar{\textbf{u}}^*_{\text{HI}}
\end{align}
\begin{align}
\label{eqn:10}
\bar{\Omega}=\bar{v}_o(\mp\bar{\mathcal{F}}&\left[\frac{3\sin2\psi}{16\bar{y}^3}+\frac{3\sin{2(\theta_c+\psi)}}{16\bar{l}^3}\right]\notag\\
&\quad
\pm\bar{\mathcal{S}}\left[\frac{3\cos\psi}{8\bar{y}^4}-\frac{3\lambda\cos{(\theta_c+\psi)}}{8(1+\lambda)\bar{l}^4}\right])
\end{align}
In Fig.~\ref{fig:S3} for the $w_2$ corner,
since we measure $\theta_c$ in the opposite direction, that of the $w_1$ corner, one needs to consider $\theta_c$ as $-\theta_c$.
In the equations, $\pm/\mp$ denotes the outcomes of $w_1$(top symbol) and $w_2$ (bottom symbol) corners. All the terms that are functions of $\bar{y}$ and $\bar{l}$ denote the terms due to the wall and interface, respectively. Here $\bar{v}_o={v_d}/{v_w}$.\\\;\
\textbf{Alternative way to solve for $w_2$ corner:} One can solve the equations of $w_1$ corner with appropriate initial conditions, later flip the sign of $\bar{v}_y$ and $\bar{\Omega}$ to get a mirror result, which is nothing but the outcomes of $w_2$ corner.

\subsection{Solution technique} We employ \textit{Wolfram Mathematica} to simplify Eqs.~\ref{eqn:4} and ~\ref{eqn:9}, yielding Eqs.~\ref{eqn:6}, ~\ref{eqn:7} and \ref{eqn:10}. Subsequently, these equations are solved using MATLAB code based on the initial-value problem with the \texttt{ODE45} solver to obtain the required unknown variables.
\subsection{Initial conditions used for generating data shown in the main text and in the supplementary material subsequently}
\noindent\textbf{Fig:3}\\
(b) $\theta_c=-25^\circ; \lambda=0; \bar{x}= 1.3; \bar{y}=2; \psi=90^\circ$\\
(d) $\theta_c=-25^\circ; \lambda=10^4; \bar{x}= 1.3; \bar{y}=2; \psi=90^\circ$\\
(e) $ \theta_c=-25^\circ;\lambda=0; \bar{x}= 4; \bar{y}=0;\psi=90^\circ$\\ 
(f) $\theta_c=-65^\circ; \bar{x}= 3.8; \bar{y}=0; \psi=90^\circ$ ($\bar{L}\approx5;\bar{l}=1.2$)\\
(g) $ \theta_c=-65^\circ; \lambda=0; \bar{x}= 2; \bar{y}=-2.4; \psi=0^\circ$\\
(h) $ \theta_c=65^\circ; \bar{x}= 2; \bar{y}=-2.8; \psi=0^\circ$\\
\newline
\textbf{Fig:4}\\
(a) and (b) $-\theta_c=25^\circ,35^\circ,45^\circ,55^\circ,65^\circ,70^\circ; \bar{l}=1\text{-}1.6;$
Approx. 5 ($\approx\bar{L}$) units away from the corner apex along the interface; $\psi=90^\circ$\\
(d) $\theta_c=-65^\circ; \bar{l}=1.2 \text{-}2; \bar{L}\approx4\text{-}7;\psi=70^\circ \text{-} 90^\circ$\\
(e) $\theta_c=-65^\circ; \bar{l}=1.6\text{-}1.7; \bar{L}\approx5;\psi=89.5^\circ$
\\
\newline
\textbf{Fig:S9}\\
(b) $\theta_c=25^\circ; \lambda=0; \bar{x}= -1; \bar{y}=-2.9; \psi=0^\circ$\\
(d) $\theta_c=25^\circ; \lambda=10^4; \bar{x}= -1; \bar{y}=-2.9; \psi=0^\circ$\\
\newline
\textbf{Fig:S10}\\
(a) $\theta_c=-25^\circ; \bar{x}= 0; \bar{y}=2.9; \psi=90^\circ$ ($\bar{L}\approx5;\bar{l}=1.2$)\\
(b)  $\theta_c=-65^\circ; \bar{x}= 3.8; \bar{y}=0; \psi=90^\circ (\bar{L}\approx5;\bar{l}=1.2)$
and $\theta_c=-25^\circ; \bar{x}= 0; \bar{y}=2.9; \psi=90^\circ$($\bar{L}\approx5;\bar{l}=1.2$)
 \\
\newline
 \textbf{Fig:S11}\\
$-\theta_c=25^\circ,35^\circ,45^\circ,55^\circ,65^\circ,70^\circ; \bar{l}=1\text{-}1.6;\bar{L}\approx5;\psi=90^\circ$\\
\newline
 \textbf{Fig:S12}\\
(a) $\theta_c=25^\circ;\lambda\rightarrow\infty;\bar{x}=0, \bar{y}=-2.8;\psi=0^\circ$\\
(b) $\theta_c=-25^\circ;\lambda\rightarrow\infty;\bar{x}=1, \bar{y}=2;\psi=90^\circ$\\
\newline
 \textbf{Fig:S13}\\ $\theta_c=25^\circ,35^\circ,45^\circ,55^\circ,65^\circ,70^\circ; \bar{x}=2; \bar{y}=-2.8;\psi=0^\circ$\\
 \newline
 \textbf{Fig:S14}\\ 
 $\theta_c=65^\circ; \bar{x}=1\text{-} 3; \bar{y}=-3\;\text{-}\;(-2);\psi=-10^\circ$-$ 10^\circ$
 
\section{NUMERICAL METHODOLOGY}
\label{sec:NM}
\subsection{Lattice Boltzmann method}
The Lattice Boltzmann method (LBM) is a numerical scheme to simulate the dynamics of fluids as an alternative to solving Navier Stokes equations directly \cite{kruger2017lattice}. In this method, a set of distribution functions $h_i (\br,\bc_i,t)$, discretised in velocity space along $\bc_i$ are defined such that the zeroth and first moments of the distribution function relate to the macroscopic quantities:
\begin{align*}
   \rho(\br,t) & = \sum_i h_i (\br,t)\\
    \rho(\br,t) \bv (\br, t) & = \sum_i \bc_i h_i (\br,t),
\end{align*}
namely, the local density and momentum of the fluid. Then the distribution functions, $h_i (\br,\bc_i,t)$ follow the discretized lattice Boltzmann equation:
\begin{align}
    h_i(\br+\bc_i \Delta t, t + \Delta t) &= h_i(\br,t)  -\notag\\
    &\quad \frac{\Delta t}{\tau}[h_i(\br,t)-h_i^{eq}(\br,t)] + \bS_i,
    \label{eqn:LBM_3}
\end{align}
where, $\bS_i$ is the discrete source (force) term, and the remaining terms on right side represents relaxation of the distributions towards a discrete equilibrium function, $h_i^{eq}$,
\begin{align}
    h_i^{eq}(\br, t) = \rho w_i \left(1 + \frac{\bv \cdot \bc_i}{c_s^2} + \frac{(\bv \cdot \bc_i)^2}{2c_s^4} - \frac{\bv \cdot \bv}{2c_s^2}\right).
    \label{LMB_4}
\end{align}
described using a BGK, single relaxation time approximation. In the above, $\rho$ is the fluid density, $w_i$ is the lattice-specific weighting factors, and $c_s = (1/\sqrt{3}) \Delta x/ \Delta t$ defines the lattice speed of sound, with $\Delta x$ and $\Delta t$ corresponding to the lattice spacing and time step, respectively. The left-hand side of Eq.~\ref{eqn:LBM_3} is a propagation step, where in the populations at $\br,t$ move to $\br+\bc_i,t+\Delta t$. Hence, computationally the algorithm follows the successive steps of collision and propagation operations.

\subsection{Diffuse Interface model}
In this work, we have used a Diffuse Interface model (DIM) to simulate the binary system of fluids. The essential feature of this model is that the interface between the two fluids is treated as a finite region with a non-zero thickness \cite{anderson1998diffuse}, which avoids tracking of the interface numerically. An order parameter field $\varphi= (n_1 - n_2)/(n_1 + n_2)$ is defined where $n_i$ being the number density of respective fluid, and $\varphi$ varies continuously between $-1$ to $1$, characterizing the two immiscible fluids. Then the evolution of the binary system is governed by \cite{zhang2022phase}:
\begin{align}
    \frac{\partial \varphi}{\partial t} + \bnabla \cdot (\varphi \bv - \mathcal{M} \bnabla \mu_o)  =  0
    \label{eqn:phi_gov}
\end{align}
where, $\bv$ is the velocity field, and $\mathcal{M}$ is the mobility. The chemical potential $\mu_o$ that governs the phase separation dynamics, is derived from the variational derivative of the free energy $\mathcal{F}(\varphi)$. For a symmetric binary fluid, $\mathcal{F}(\varphi)$ is expressed as \cite{kendon2001inertial}:
\begin{align}
    \mathcal{F}(\varphi) &= \frac{1}{2}A\varphi^2 + \frac{1}{4} B \varphi^4 + \frac{1}{2} \mathcal{C} |\nabla \varphi|^2, 
\end{align}
where $A$ and $B$ are phenomenological constants and we choose $B=-A$, $\mathcal{C}$ corresponds to an increase in the free energy of the system due to gradients in the order parameter.
For a one-dimensional interface, say along the x-axis, the order parameter field at equilibrium is given by,
\begin{align}
        \varphi &=- \varphi^* \tanh \left(\frac{x}{\xi}\right),
    \label{eqn:PE_8}
\end{align}
where $\varphi^* = \sqrt{-A/B} = \pm 1$, the order parameter field in the bulk of the fluid, and $\xi= (-2k/A)^{1/2}$ denotes the interfacial thickness. The corresponding interfacial tension ($\sigma$), given by the excess free energy per unit interfacial area, is obtained from:
\begin{align*}
    \sigma &= \int_{-\infty}^\infty \left[\frac{\mathcal{C}}{2} \left( \frac{d \varphi}{dx}\right)^2 + \mathcal{F}(\varphi) - \mathcal{F}(\varphi_{\text{bulk}})\right] dx, \\
    & = \frac{4\mathcal{C} \varphi^{*2}}{3\xi}
\end{align*}
In this work, we employ a finite difference method to solve Eqn.~\ref{eqn:phi_gov}.  Initially, the advective fluxes are computed based on the velocity field and the order parameter field. Subsequently, the diffusive fluxes are determined using the chemical potential field. Boundary conditions, if applicable, are then enforced on the fluxes. Finally, the order parameter field is updated by incorporating the computed fluxes.
\subsection{Squirmer model}
In this study, we use the squirmer model to simulate the behaviour of the microdroplet. The squirmer model, introduced by Lighthill and Blake \cite{blake1971spherical}, represents a non-deformable sphere of radius $R_d$, propelled by a prescribed tangential slip velocity  $\bu_{\text{slip}} (\psi)$ on its surface.
Assuming that the surface slip is purely in the tangential direction, $\be_\psi$, the slip velocity is given by
\begin{align}
    \bu_{\text{slip}}(\psi)  = B_1( \sin \psi + \beta \sin \psi \cos \psi) \be_\psi
    \label{eqn:sq_2}
\end{align}
where, $\beta (=  B_2/B_1)$ is a dimensionless parameter that characterizes the squirmer strength and its type. Specifically, a positive value of $\beta$ corresponds to a puller, a negative value indicates a pusher, while $\beta = 0$ characterizes a neutral squirmer.

\subsection{Combining LBM, DIM, and the squirmer models}
In the above, we introduced individual solution methodologies/models: LBM to solve the fluid flows, DIM to model the binary fluid system, and the squirmer to model the droplet. All three models are combined to computationally simulate the experimental system. The fluid motion is generated by the imposed boundary conditions on the surface of the squirmer as well as the interfacial stresses generated in the DIM. The dynamics of the squirmer are simultaneously influenced by the stresses exerted by the fluid.

Considering the translational velocity $\bar{\bV}$, and rotational velocity $\bar{\mathbf{\Omega}}$ of the squirmer, the surface velocity $\bv_s$ is expressed as \cite{purushothaman2021hydrodynamic}: 
\begin{align}
    \bv_s = \bar{\bV} + \bar{\mathbf{\Omega}} \times (\br_s- \br_o) + \bv_{\text{slip}},
    \label{eqn:sq_3}
\end{align}
where, $\br_s, \br_o$ represent the position vectors of a point on the surface and the centroid of the squirmer, respectively. The surrounding fluid dynamics are thus governed by Eqn. \ref{eqn:sq_3} and are solved using LBM. 

Now, to determine the effect of fluid flow on the squirmer’s motion, the surface force distribution is obtained from the first moment of the distribution function. The local force per unit area acting on a surface element is given by:
\begin{equation}
    \bf_s = \sum_i -2\bc_i \left[h_i(\br, t^+) + \frac{\rho w_i}{c_s^2} (\bv_s \cdot \bc_i)\right]
    \label{eqn:sq_4}
\end{equation}
where, $t^+$ denotes the post-collision time and $\bf_s$  is the surface traction on the squirmer. The total force $\bF_s$ and torque $\bT_s$  acting on the squirmer are then obtained by integrating the force distribution over its surface:
\begin{align}
    \int \bf_s dS &= \bF_s,\\
    \int (\br_s- \br_o) \times \bf_s dS &= \bT_s.
\end{align}
These hydrodynamic quantities are subsequently used to update the linear and angular velocities of the squirmer through its equations of motion:
\begin{align}
    m_s \bar{\bV} (t + \Delta t) &= m_s \bar{\bV} (t) + \Delta t \bF_s(t)\\
    I_s \bar{\mathbf{\Omega}}(t + \Delta t) &= I_s \bar{\mathbf{\Omega}}(t) + \Delta t \bT_s(t),
\end{align}
where $m_s$ and $I_s$ denote the mass and moment of inertia of the microswimmer, respectively.

It may noted that, interfacial stresses generated by DIM model appears as a force contribution in LBM, and hence the dynamics due to interfacial stresses are automatically accounted for while LBM is performed. The generated fluid flows advect the order parameter field, and thus describe any interfacial deformations. In our simulations, capillary number (ratio of viscous to surface tension forces) is kept smaller so that interfacial deformation, and its dynamics are not significant.

\subsection{Simulation details}
\begin{figure}[h!]
  
  \centerline{\includegraphics[width=8cm]{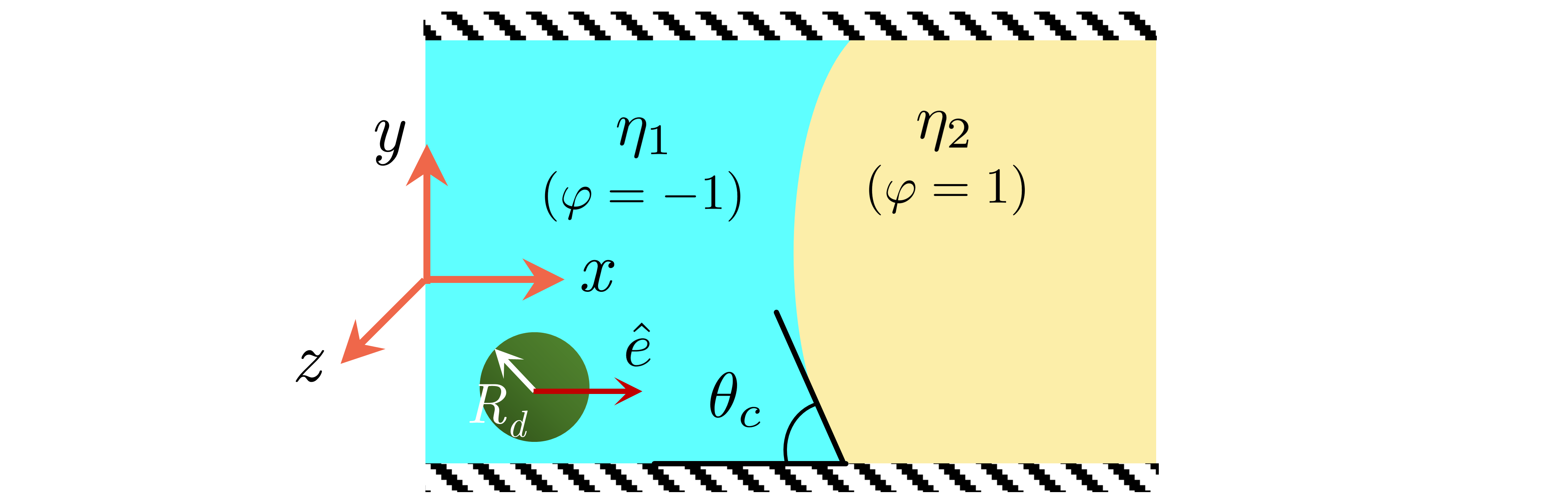}}
  \caption{Schematic illustration of a microswimmer of radius $R_d$ confined in a rectangular channel adjacent to a curved fluid-fluid interface, initialized parallel to the $x$-axis near the bottom wall. The two immiscible fluids, represented in blue and yellow, have viscosities $\eta_1$ and $\eta_2$, respectively. The black dashed line marks the interface, with $\theta_c$ denoting the contact angle at the wall. The red arrow indicates the microswimmer's orientation and direction of propulsion with velocity $\bar{V}$. The thick black lines denote the no-slip boundaries of the channel.}
\label{fig:S4}
\end{figure}
In this work, simulations are carried out in a computational domain of $256 \times 80 \times 28$. Both the lattice spacing $\Delta x$ and time increment $\Delta t$ are set to $1$.  Unless otherwise specified, all the quantities are reported in lattice Boltzmann units but appropriately non-dimensionalised to compare with experimental/theoretical results. The walls are modelled with no-slip boundary conditions (i.e., in the $y$ and $z$ direction), while periodic boundary conditions are enforced in directions without confining walls (i.e., in the $x$-direction). The fluid density is uniformly set to $1$. The dynamic viscosity varies spatially based on the phase field variable $\varphi$, following an Arrhenius-type relation \cite{langaas2000lattice}:
\begin{align*}
    \eta_{\varphi} &= \eta_1 ^{\left(\frac{1 - \varphi}{2}\right)} \eta_2 ^{\left(\frac{1 + \varphi}{2}\right)},
\end{align*}
where, $\eta_{\varphi = -1} = \eta_1~ \text{and} ~\eta_{\varphi = 1} = \eta_2$. Each fluid occupies half of the domain, positioning the fluid-fluid interface at $x = 128$. The squirmer has a diameter of $10$ LB units. Each simulation runs for a minimum duration of $1200000 \Delta t$, with data snapshots stored every $2000\Delta t$ for subsequent analysis. To account for the spurious velocities generated near the interface in the lattice Boltzmann simulations, the reported flow fields are calculated by subtracting the velocity field generated solely by the interface from that obtained when both the swimmer and the interface are present, for a particular time step. 

\section{SUPPLEMENTARY RESULTS}
\label{sec:SR}
\begin{figure}[H]
\centerline{\includegraphics[width=8cm]{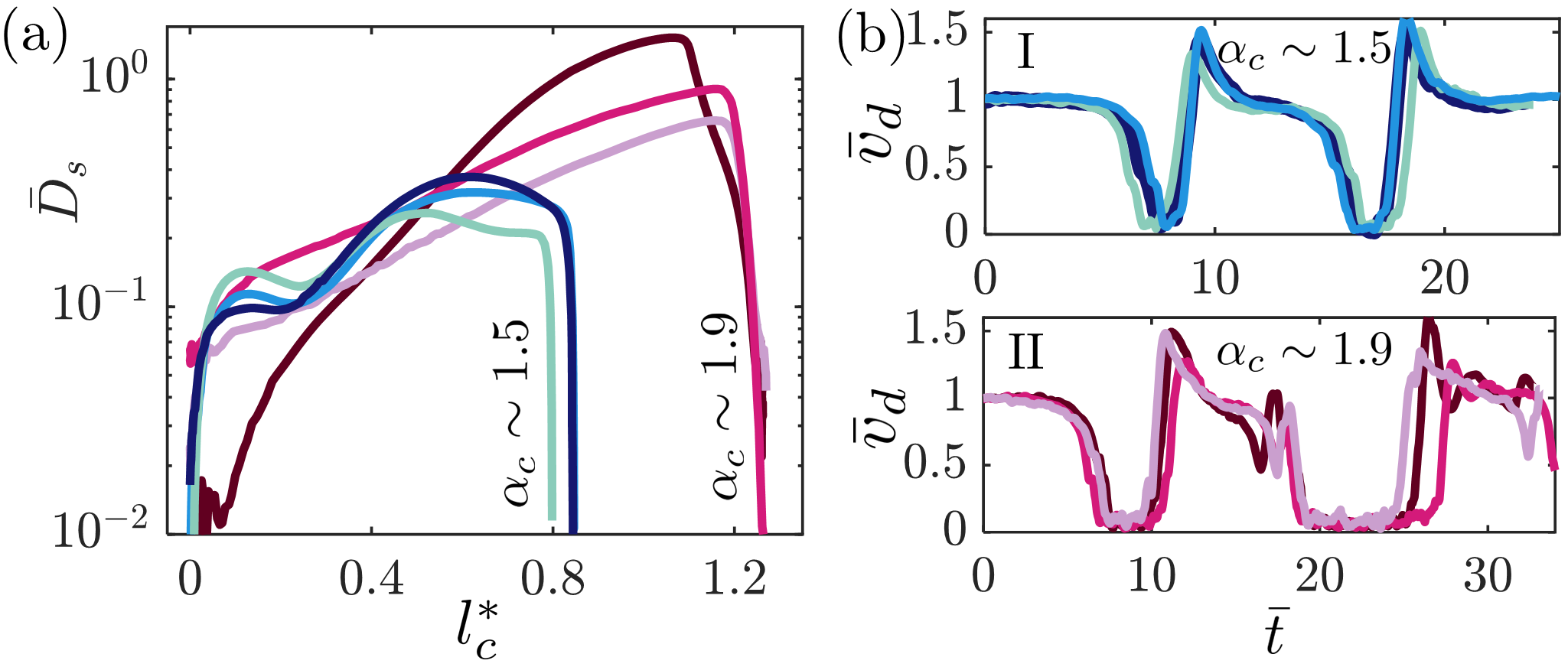}}
  \caption{(a) Variations of the non-dimensional separation distance ($\bar{D}_s$) between the microswimmer and the air-water meniscus along the non-dimensional curvilinear coordinate ${l^*_c}$ (see Fig.~\ref{fgr:1}(b)) for increasing values of the non-dimensional parameter $\alpha_c$ for multiple experimental outcomes.
 (b) Temporal variations of $\bar{v}_d$ as the microswimmer approaches, swims along, and leaves the air-water meniscus for increasing value of $\alpha_c$ (I-II) for multiple cases to show the consistency of the results.}
\label{fig:S5}
\end{figure}

\begin{figure}[H]
\centerline{\includegraphics[width=9cm]{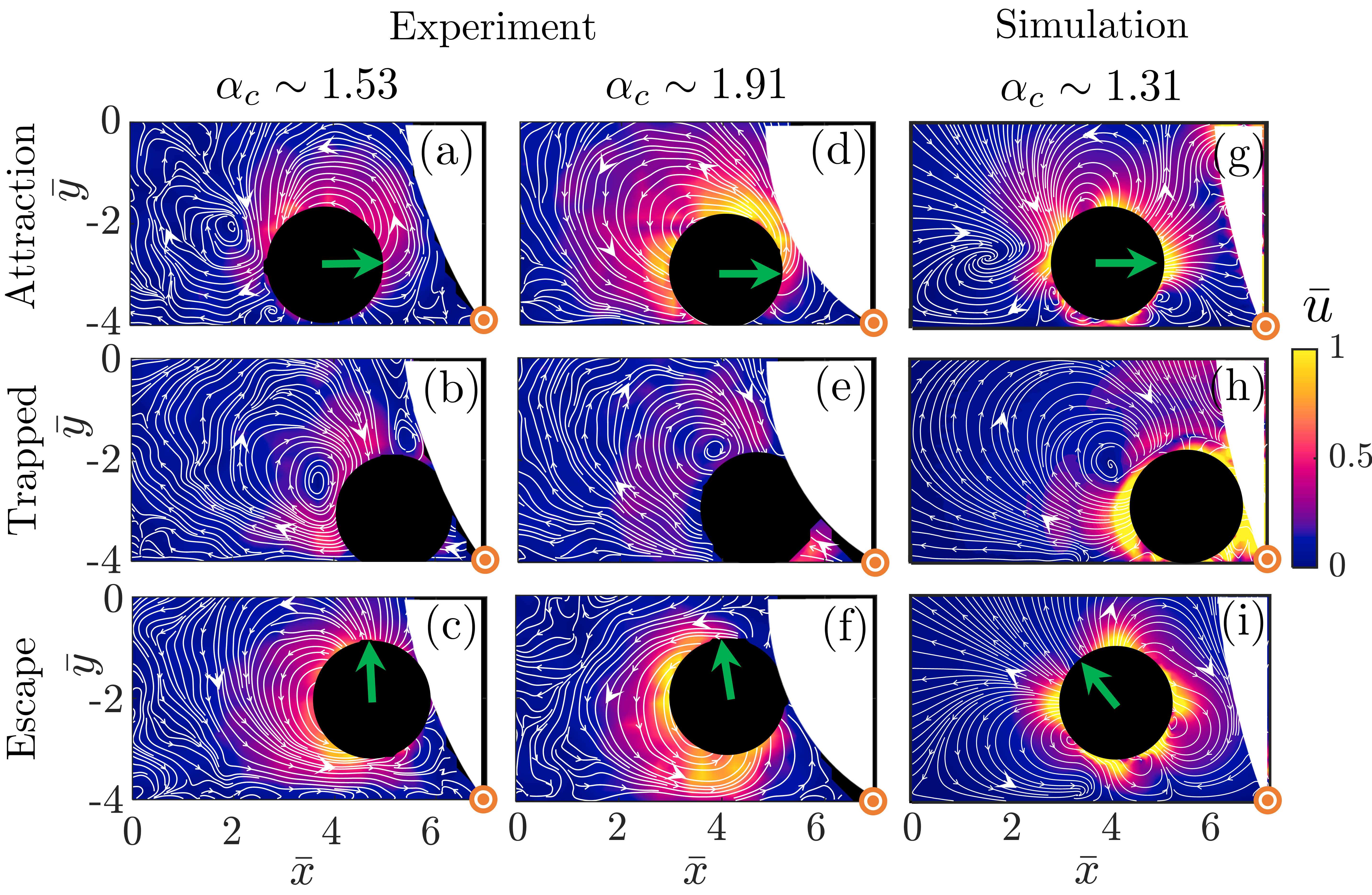}}
  \caption{Variations in the flow field generated by the microswimmer near the corner of a meniscus in a microchannel. Changes in the hydrodynamic signature of the active droplet (represented by streamlines and local flow speed $\bar{u}$ contour plots) as it gets attracted to, trapped at, and escapes from the corner ($w_1$) of an air-water meniscus for (a)-(c) $\alpha_c = 1.53$ ($\theta_c=55^\circ$), and (d)-(f) $\alpha_c = 1.91$ ($\theta_c=25^\circ$). (g)-(i) Corresponding flow fields obtained from the Lattice Boltzmann Method (LBM) based numerical simulations of a model, self-propelling microswimmer (squirmer; $\beta=-1$) near a corner ($\alpha_c = 1.31$; $\theta_c=65^\circ$) of a confined meniscus similar to that in (a)-(c).
  }
\label{fig:S7}
\end{figure}

\begin{figure}[H]
  \centerline{\includegraphics[width=8cm]{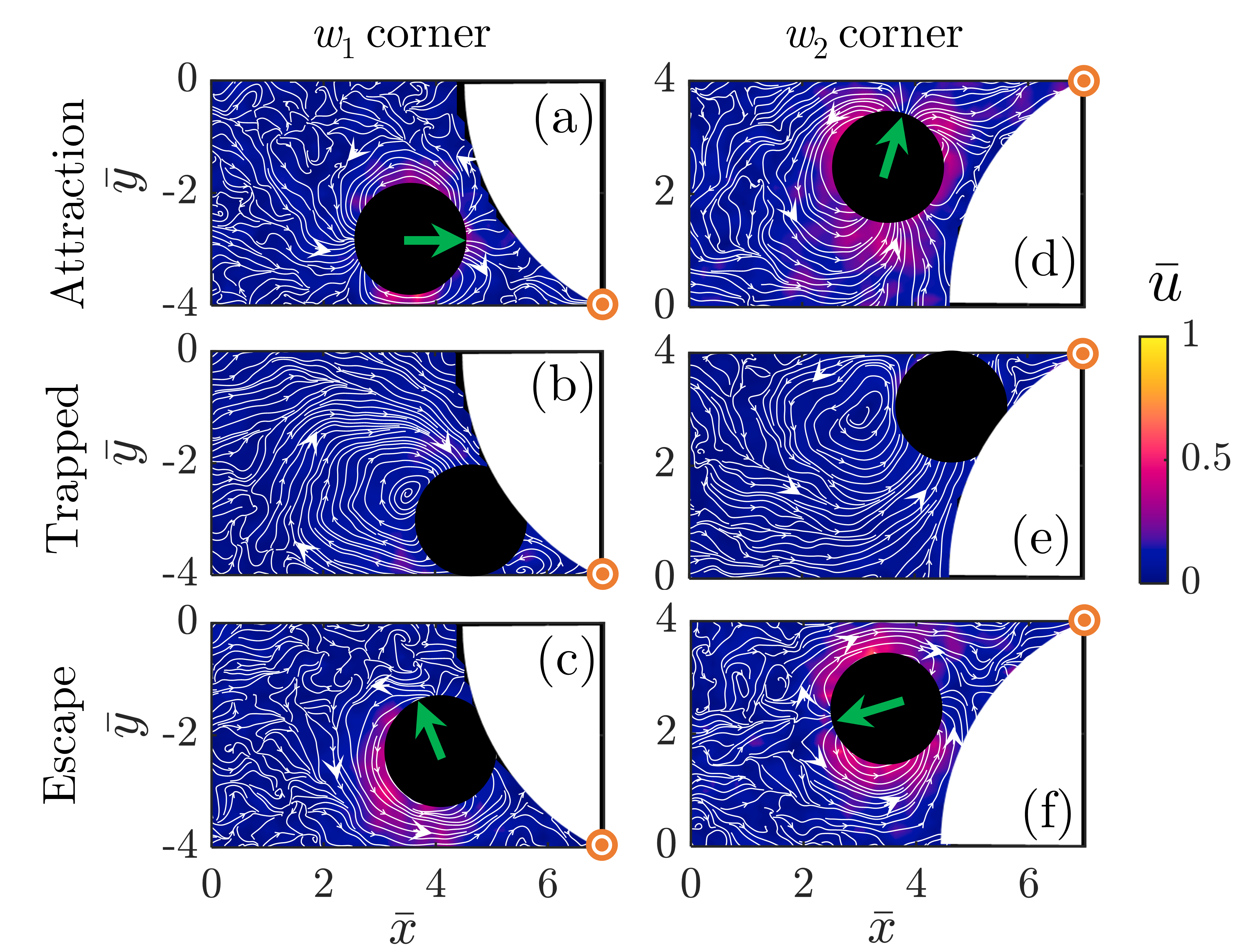}}
  \caption{Variations in the flow field generated by the microswimmer near the corner of an oil-water ($\lambda=10^4$) meniscus in a microchannel. Changes in the hydrodynamic signature of the active droplet as it gets attracted to, trapped at, and escapes from the corner ($w_1$) for (a)-(c), and (d)-(f)) for $w_2$ corner ($\alpha_c = 1.96;\;\theta_c=25^\circ$).}
\label{fig:S8}
\end{figure}

\begin{figure}[H]
  \centerline{\includegraphics[width=8.6cm]{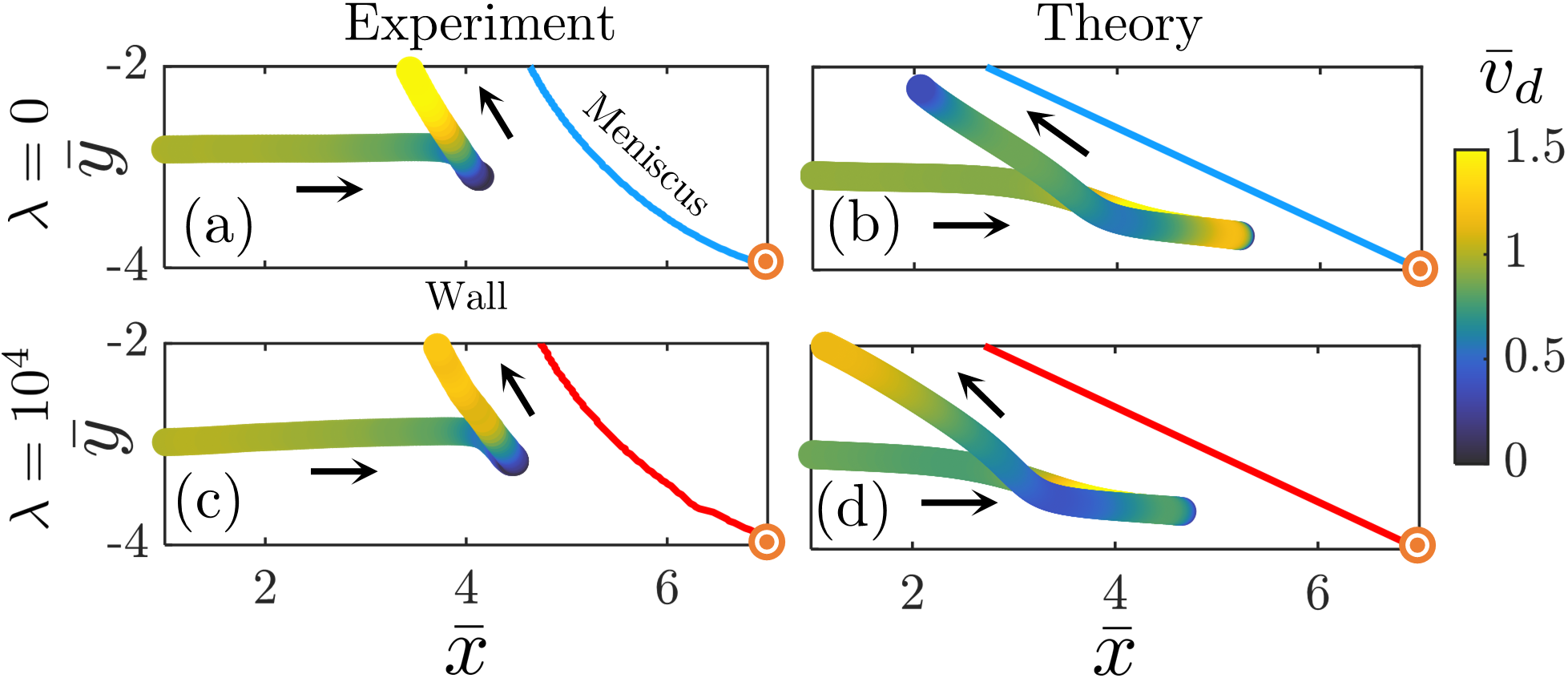}}
  \caption{Comparison between experimental ((a),(c)) and theoretical ((b),(d)) swimming trajectories for the microswimmer (squirmer; $\beta=-1$) near the $w_1$ corner $(\theta_c \approx 25^\circ; \alpha_c=1.90)$  of a confined air-water meniscus $(\lambda=0)$ and an oil-water meniscus $(\lambda=10^4)$.}
\label{fig:S9}
\end{figure}

\begin{figure}[H]
  \centerline{\includegraphics[width=9cm]{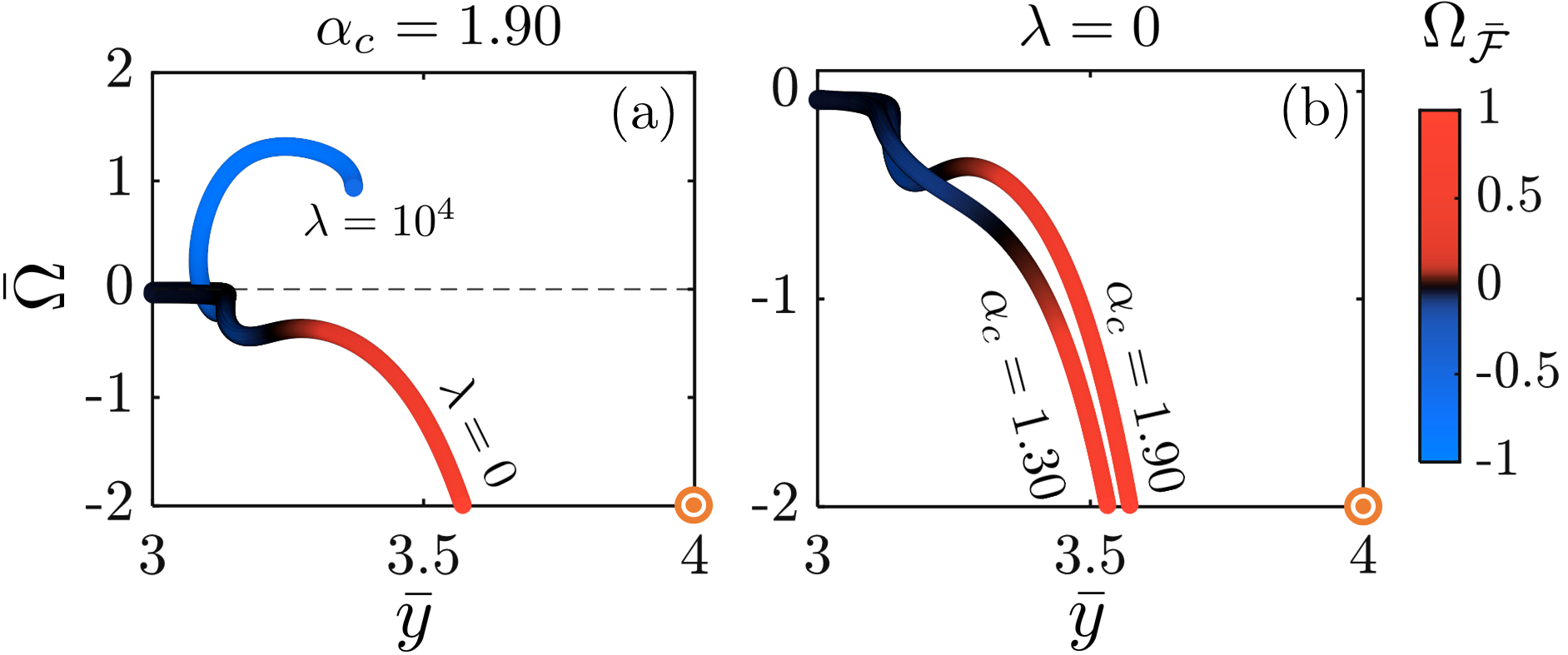}}
  \caption{Theoretical variations of rotational swimming velocity  $\bar{\Omega}$ with $\bar{y}$ for (a) different $\lambda\;(\alpha_c=1.90)$ and (b) different $\alpha_c\;(\lambda=0)$ near the $w_2$ corner;
$\bar{\Omega}$ variations are colour-coded with the contribution from the force dipole term $(\bar{\Omega}_{\bar{\mathcal{F}}})$. }
\label{fig:S10}
\end{figure}

\begin{figure}[H]
  \centerline{\includegraphics[width=9cm]{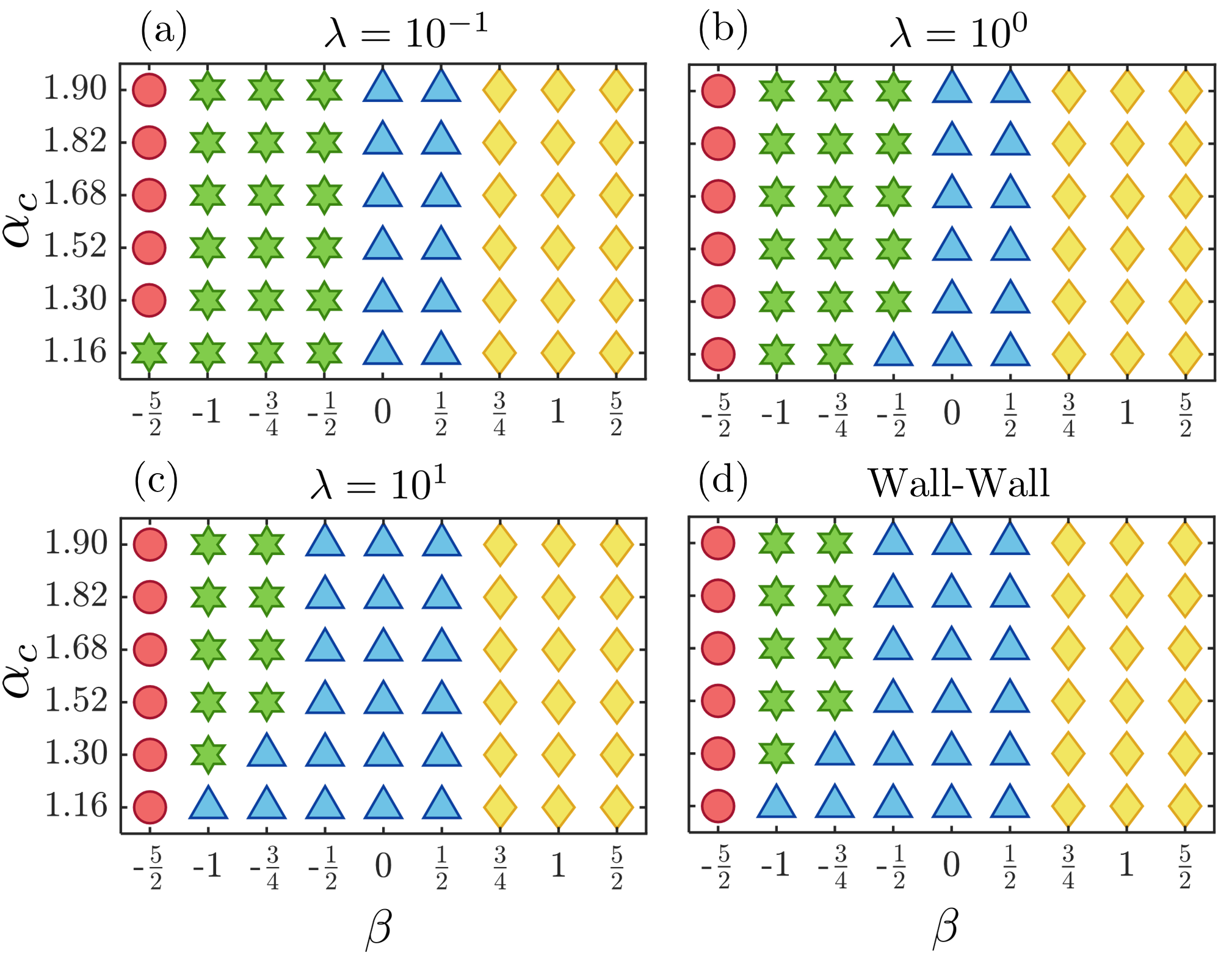}}
  \caption{Shape of the top $(w_2)$ corner $(\alpha_c)$, microswimmer strength $(\beta)$ state diagrams for (a) $\lambda=10^{-1}$, (b) $\lambda=10^0$, (c) $\lambda=10^{1}$, and (d) corner made of two rigid walls.}
\label{fig:S11}
\end{figure}

\begin{figure}[H]
  \centerline{\includegraphics[width=8.5cm]{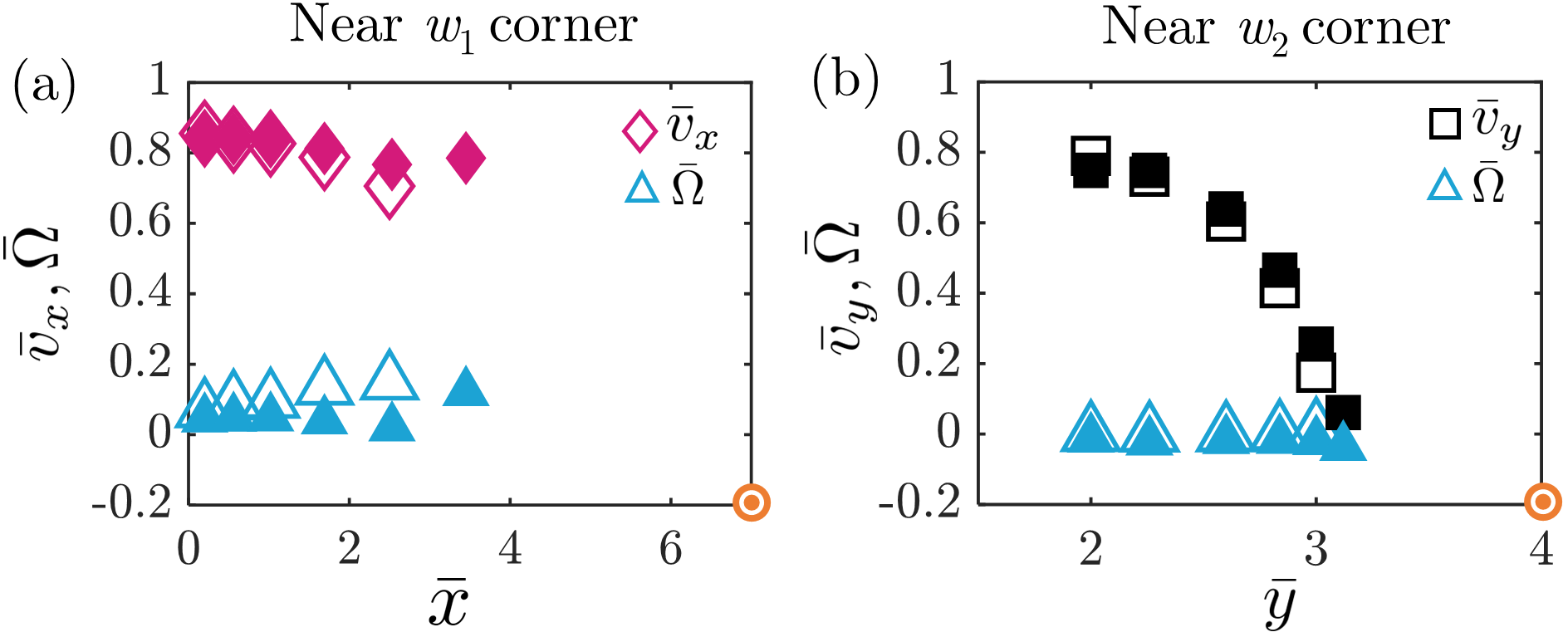}}
  \caption{Variations in (a) 
  axial velocity $\bar{v}_x$ and $\bar{\Omega}$ with axial position $(\bar{x})$ as the microswimmer approaches the $w_1$ corner $(\theta_c= 25^\circ; \alpha_c=1.90)$ for a solid-solid corner and in (b) the transverse velocity component $(\bar{v}_y)$ and angular velocity $(\bar{\Omega})$ with the transverse location $(\bar{y})$ as the microswimmer approaches the $w_2$ corner; filled and hollow markers are from theory and hybrid lattice Boltzmann simulations, respectively.}
\label{fig:S12}
\end{figure}

\begin{figure}[H]
  \centerline{\includegraphics[width=9cm]{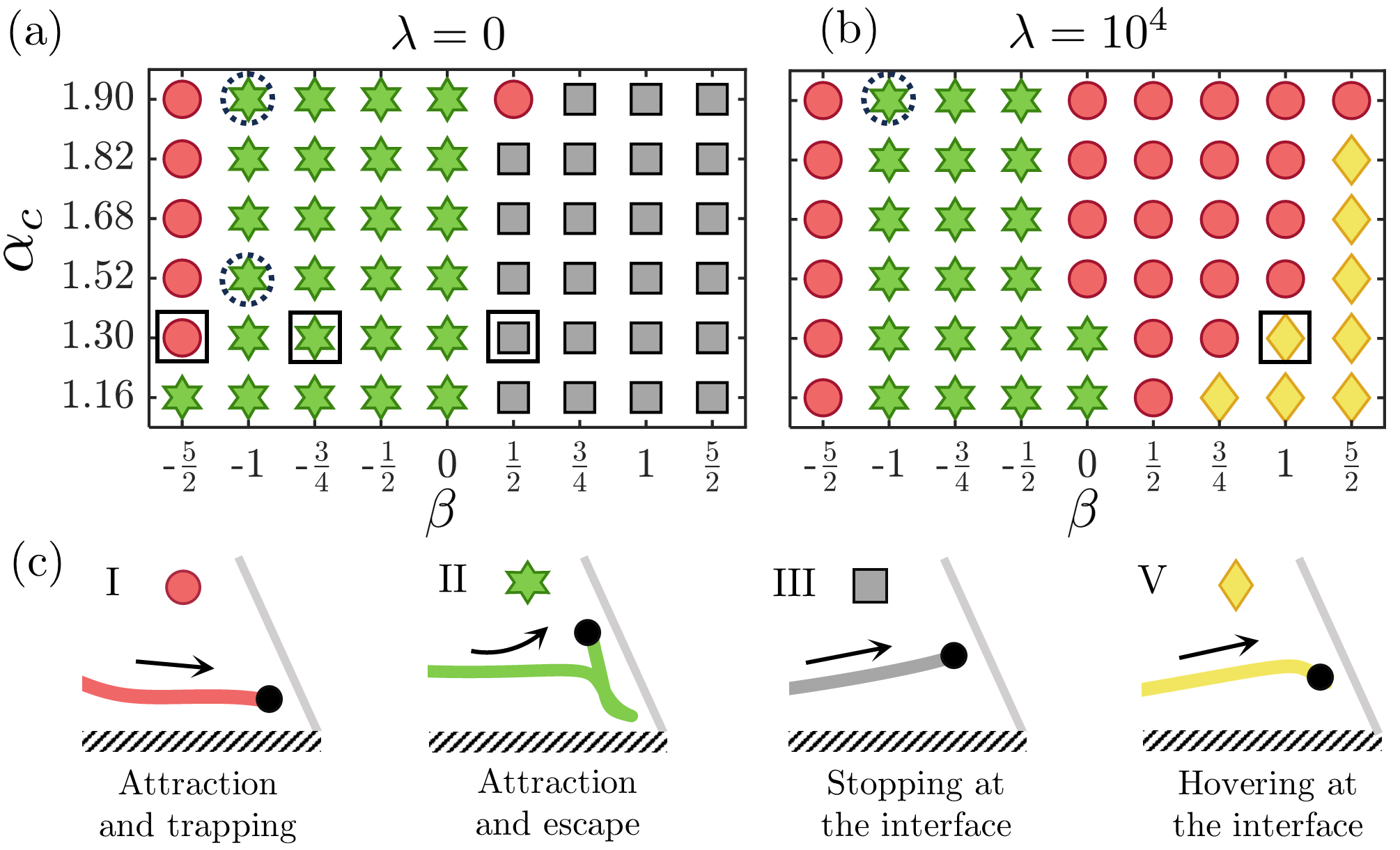}}
  \caption{Shape of the bottom $(w_1)$ corner $(\alpha_c)$, microswimmer strength $(\beta)$ state diagrams for (a) an air-water meniscus $(\lambda=0)$, and (b) an oil-water meniscus $(\lambda=10^4)$.
 The symbols correspond to the 4 different types of swimmer dynamics as explained in (c) for the respective parameter, as in the respective same-colour square box. The dotted circle represents the comparable $\alpha_c$ with experimental outcomes (Figs.~\ref{fig:S9}(a), (c).}
\label{fig:S13}
\end{figure}

\begin{figure}[H]
  \centerline{\includegraphics[width=9cm]{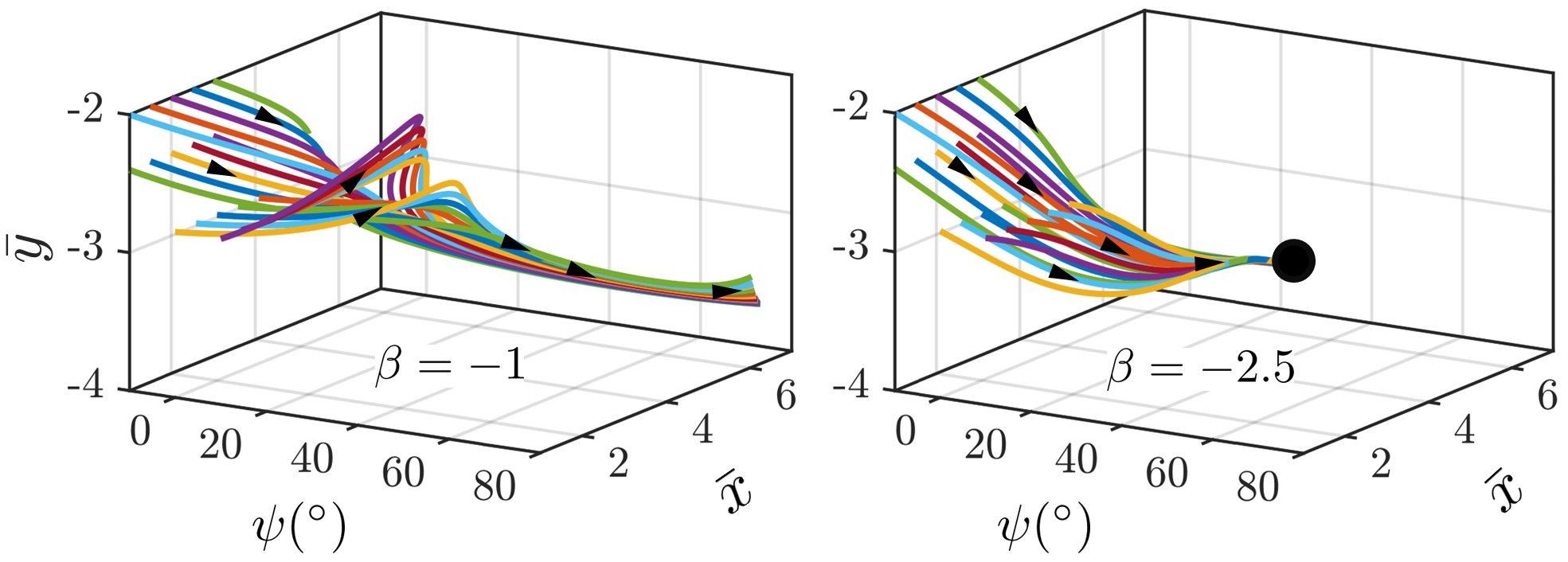}}
  \caption{Microswimmer trajectories in $x-y-\psi$ plane, near the $w_1$ corner of an air-water interface ($\alpha=1.30$) with different initial conditions, show that the vicinity of the three-phase contact line (TPCL) acts like a fixed point which changes from unstable to stable with reducing $\beta$.}
\label{fig:S14}
\end{figure}

\section{VIDEOS OF MICROSWIMMER NEAR
MENISCUS CORNER}
\label{sec:Videos}
All the videos are provided with real timestamps.
\begin{figure}[H]
\centerline{\includegraphics[width=6cm]{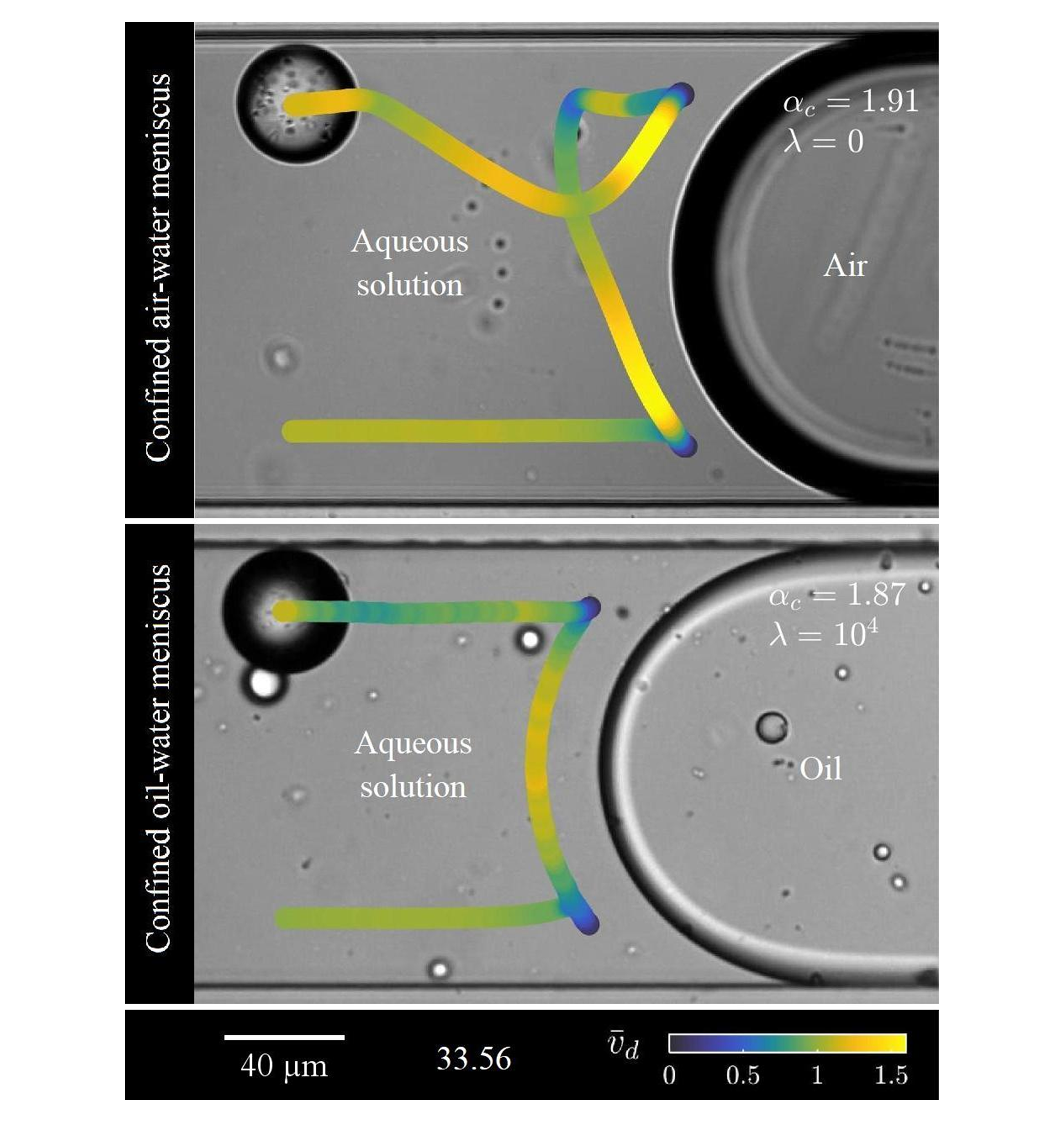}}
Bright-field microscopy captured footage of droplet microswimmers near the air-water (top) and oil-water (bottom) menisci inside the microchannel. For video, click $\rightarrow$\href{https://drive.google.com/file/d/1YUDQ6YEPD92A4A1s1wlKv4-hXzRiWWpU/view?usp=sharing}{\textbf{Video S1}}.
\label{fig:Video-S1}
\end{figure}

\begin{figure}[H]
\centerline{\includegraphics[width=8cm]{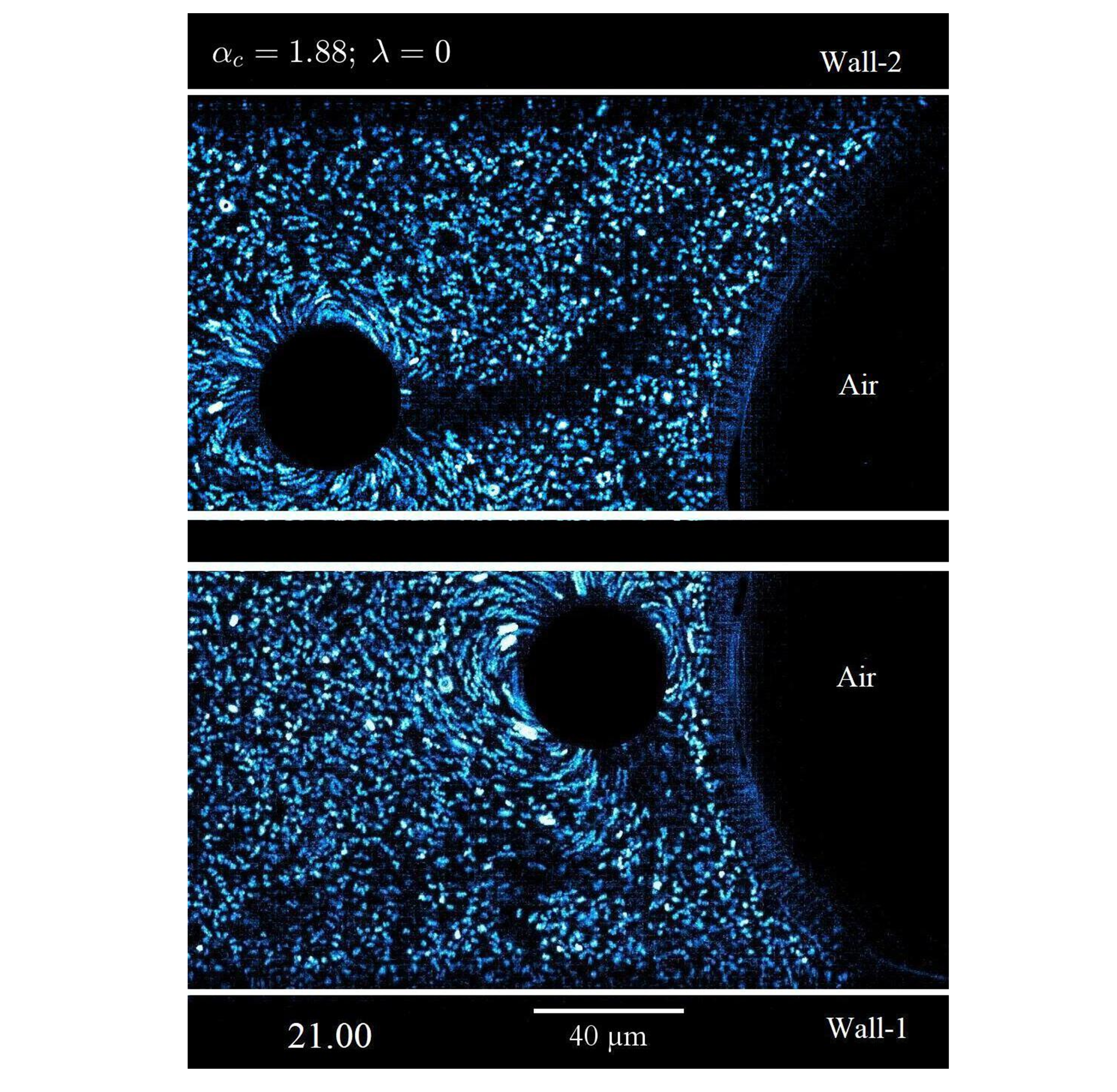}}
Video of an active droplet near the air-water meniscus. For video, click $\rightarrow$ \href{https://drive.google.com/file/d/1E0tsuMvlyc87mvgpSJlxDkFGiHuxJJEG/view?usp=drive_link}{\textbf{Video S2}}. Video editing is performed using ImageJ \cite{ImageJ} with the FlowTrace \cite{Flow_Trace, Flow_Trace_2} extension. 
\label{fig:Video-S2}
\end{figure}

\begin{figure}[H]
\centerline{\includegraphics[width=8cm]{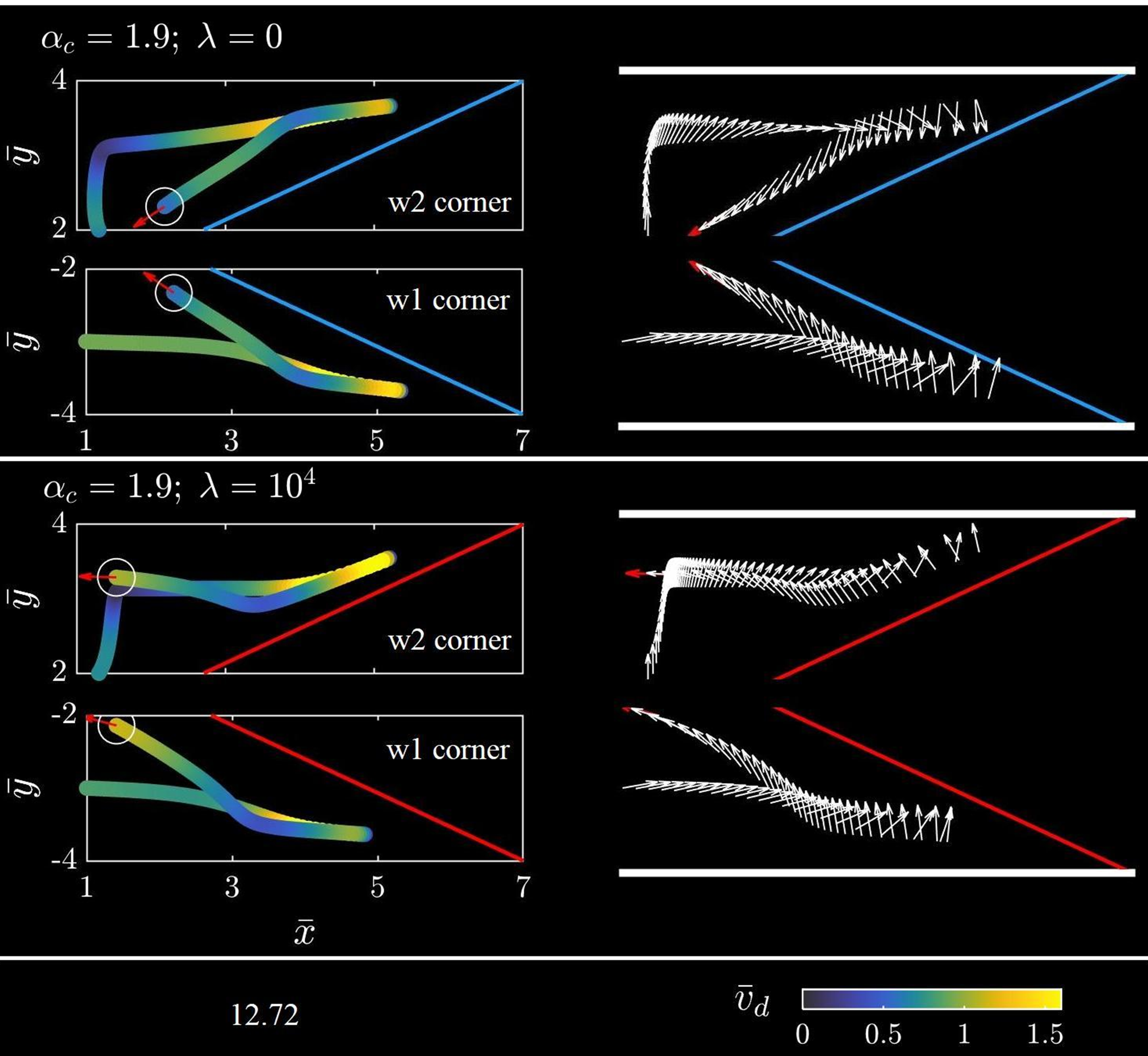}}
Videos of the microswimmer near air-water ($\lambda=0$) and oil-water ($\lambda=10^4$) are displayed, with trajectories colour-coded by average terminal speed and the corresponding head orientation shown in the adjacent column. For video click $\rightarrow$ \href{https://drive.google.com/file/d/1r8NkvrolyOQG-ujNu3LqQNP6JPlU6SR6/view?usp=drive_link}{\textbf{Video S3}}. 
\label{fig:Video-S3}
\end{figure}

\end{document}

%% file: definitions.tex
\def \bnabla{\boldsymbol{\nabla}}

\def \bF{\mathbf{F}}

\def \bS{\mathbf{S}}

\def \bT{\mathbf{T}}

\def \bV{\mathbf{V}}

\def \bc{\mathbf{c}}

\def \be{\mathbf{e}}

\def \bf{\mathbf{f}}

\def \br{\mathbf{r}}

\def \bu{\mathbf{u}}

\def \bv{\mathbf{v}}